\documentclass{jppHack}

\usepackage{amsmath}
\usepackage{graphicx}
\usepackage[usenames]{color}

\newcommand{\ApJ}{\mbox{\em Astrophys. J. }}
\newcommand{\ApJSupp}{{\em Astrophys. J. Supp. }}
\newcommand{\AandA}{\mbox{\em Astron. Astrophys. }} 
\newcommand{\MNRAS}{{\em Monthly Notices Royal Ast. Soc. }}

\ifCUPmtlplainloaded \else
  \checkfont{eurm10}
  \iffontfound
    \IfFileExists{upmath.sty}
      {\typeout{^^JFound AMS Euler Roman fonts on the system,
                   using the 'upmath' package.^^J}%
       \usepackage{upmath}}
      {\typeout{^^JFound AMS Euler Roman fonts on the system, but you
                   dont seem to have the}%
       \typeout{'upmath' package installed. JPP.cls can take advantage
                 of these fonts, if you use 'upmath' package.^^J}%
      }
  \else
  \fi
\fi

\ifCUPmtlplainloaded \else
  \checkfont{msam10}
  \iffontfound
    \IfFileExists{amssymb.sty}
      {\typeout{^^JFound AMS Symbol fonts on the system, using the
                'amssymb' package.^^J}%
       \usepackage{amssymb}%

      }{}
  \fi
\fi

\ifCUPmtlplainloaded \else
  \IfFileExists{amsbsy.sty}
    {\typeout{^^JFound the 'amsbsy' package on the system, using it.^^J}%
     \usepackage{amsbsy}}
    {}
\fi

\newsavebox{\astrutbox}
\sbox{\astrutbox}{\rule[-5pt]{0pt}{20pt}}

\newcommand\etal{\mbox{{et al. }}}

\def \gtw{\>\hbox{\lower.25em\hbox{$\buildrel >\over\sim$}}\>}
\def \ltw{\>\hbox{\lower.25em\hbox{$\buildrel <\over\sim$}}\>}

\def \be{\begin{equation}}
\def \ee{\end{equation}}
\def \bold#1{\mathbf{#1}}

\def \om{\omega}
\def \g{\gamma}

\def \omp{\omega_p}

\def \pccm3{pc-cm$^{-3}$}

\title[Radio Emission Physics in the Crab Pulsar]
{Radio Emission Physics in the Crab Pulsar}

\author[J. A. Eilek and T. H. Hankins]
{J\ls E\ls A\ls N\ns A.\ns E\ls I\ls L\ls E\ls K$^{1,2}$%
  \thanks{Email address for correspondence: jeilek@aoc.nrao.edu},\ns
\and T\ls I\ls M\ls O\ls T\ls H\ls Y\ns 
 H.\ns H\ls A\ls N\ls K\ls I\ls N\ls S$^{1,2}$}

\affiliation{$^1$Physics Department, New Mexico Tech, Socorro NM 87801 USA
\\[\affilskip]
$^2$National Radio Astronomy Observatory, Socorro NM 87801 USA}

\pubyear{2010}
\volume{650}
\pagerange{119--126}
\date{?; revised ?; accepted ?. - To be entered by editorial office}
\begin{document}

\maketitle

\begin{abstract} 
We review our high-time-resolution radio observations of the Crab
pulsar and compare our data to a variety of models for the emission
physics. The Main Pulse and the Low-Frequency Interpulse come from
regions somewhere in the high-altitude emission zones (caustics) that also
produce pulsed X-ray and $\g$-ray emission.  Although no emission
model can fully explain these two components, the most likely models
suggest they arise from a combination of beam-driven
instabilities, coherent charge bunching and strong electromagnetic turbulence.
Because the radio power fluctuates on a wide range of timescales, we know the
emission zones are patchy and dynamic.  It is 
tempting to invoke unsteady pair creation in high-altitude gaps as
source of the variability, but current pair cascade models cannot explain
the densities required by any of the likely models.  It is harder to account
for the mysterious High-Frequency Interpulse.   We  understand
neither its origin within the magnetosphere  nor the striking emission
bands in its dynamic spectrum.  The most promising models are based on
analogies with solar zebra bands, but they require unusual plasma structures
which are not part of our standard picture of the magnetosphere. 
We argue that 
radio observations can reveal much about the upper magnetosphere, but 
work is required before the models can address all of the data.

\end{abstract}


\section{Introduction}

High-energy emission in pulsars comes from high altitudes.  Recent
work shows that most optical, X-ray and $\gamma$-ray
emission comes from caustic zones at mid-to-high altitudes in the
star's magnetosphere or wind.  The details, however, remain
unclear. The high-energy\footnote{We use
  ``high-energy'' to refer to optical, X-ray and $\gamma$-ray bands;
  as distinct from the radio band which is our focus in this paper.}
 emission zones
can lie anywhere along or within the open field line zones which exist
over the star's magnetic poles.  They may also coincide with current
sheets either within or outside of the light cylinder.  The highly
nonlinear behavior of relativistic photon caustics, combined with
uncertainties in the magnetic field structure close to the light
cylinder, means that any of these possible emission zones can
reproduce observed high-energy light curves. Because we do
not know whether pulsed high-energy emission is due to synchrotron
radiation, curvature radiation or inverse Compton scattering, we have
few constraints on physical conditions within the high-energy emission
zones. We need more information, and  believe radio observations can
help. 

Radio emission in a large number of  pulsars comes from
lower altitudes, close to
the magnetic axis. This geometry works for  many pulsars, but
fails for an interesting minority.  For those stars -- including the pulsar
in the Crab Nebula, the subject of this paper -- the similarity of the radio
and high-energy light curves strongly suggests the radio emission comes from
the same high-altitude regions that create the high-energy emission.  However,
 the geometrical models do not address the physical mechanisms by which
the magnetospheric plasma produces the  intense
radio emission we observe.  Many different emission mechanisms have been
proposed, but none has been proven to be operating in the general pulsar
population. 

Our group has approached the question of pulsar radio emission
mechanisms by carrying out detailed studies of one object, the Crab
pulsar.  This pulsar shines in radio at many phases
throughout its rotation period,
with seven components  detectable in its mean profile\footnote{We
  use the term ``mean profile'' to describe the radio intensity as a
  function of rotation phase.  This quantity is derived by summing the
  received intensity synchronously with the star's rotation period.
  The same data product is often called the ``light curve'',
  especially in the high-energy community.}.  At least three, and
probably five, of these come from high altitude emission zones,
spatially coincident with the high-energy emission zones in this
star's magnetosphere.  The high time resolution and broad spectral
bandwidth of our observations show that different components have
different temporal and spectral characteristics. We infer that more than
one type of radio emission mechanism is taking place
 within the star's magnetosphere.

In this paper we collect and review our observational results, compare
them to competing radio emission models, and from this comparison
discuss what physical conditions exist in the high-altitude zones
that emit both radio and high-energy pulsed emission. We begin in
Sec. \ref{EmissionZonesHighandLow} by reviewing  the basic picture
of the rotating pulsar magnetosphere and its modern extension to the
high-altitude transition to the wind zone.  We focus on
where in the extended magnetosphere the radio and high-energy emission
originate.  In Sec. \ref{HowDoTheyShine} we lay out physical
conditions  that must exist (the ``emission physics'') if the
magnetosphere is to produce intense radio emission.  In
Sec. \ref{CrabInRadio} we set the stage for our discussion of the
radio properties of the Crab.  We present the different components of
the mean radio profile and review our observations of the pulsar.  In
Sec. \ref{MainPulseBursts} we focus on the Main Pulse and the
Low-Frequency Interpulse, which we believe involve similar emission
physics.  Both of these components show bursting behavior on
sub-microsecond timescales; we argue this reflects variability of the
driving mechanism, which is probably unshielded electric fields local
to the region.  We also introduce nanoshots -- flares of emission on
nanosecond timescales -- which we believe reveal the fundamental
emission process in these components.  In Sec. \ref{Nanoshot_models}
we compare observed nanoshot properties to predictions of three
plausible radio emission models.  Although none of the models can
fully explain the data, we suggest they can be used to constrain
physical conditions in the radio emission zones. We then switch to
the High-Frequency Interpulse, a separate component with
different radio characteristics that very likely involves different 
emission physics.  In Sec. \ref{HFIP} we discuss its radio properties,
including the unexpected spectral emission bands, and consider what
emission physics might be responsible for the bands.  In Sec.
\ref{HFIP_environment} we present two additional clues to its local
environment:  signal dispersion and polarization angle. Finally,
in Sec. \ref{SomethingAtTheEnd}, after congratulating the 
steadfast reader who has made it to the end, we end by summarizing the
lessons we have learned to help us going forward.

We relegate to the Appendices necessary details on a variety of possible
radio emission mechanisms.  In App. \ref{App:RadioEmissionModels} we
briefly review specific models of coherent radio emission that have
been proposed for the general pulsar population, and in
App. \ref{App:DPR} we review one additional model specifically
proposed for the High-Frequency Interpulse of the Crab pulsar.

\section{Emission zones in the pulsar magnetosphere}
\label{EmissionZonesHighandLow}

Consider a rapidly rotating neutron star which
supports a strong magnetic 
field\footnote{The surface magnetic field strength is inferred 
from the rate at which star's rotation slows down.  The loss of
rotational energy is ascribed to magnetic dipole radiation.  
Typical values are  $B_* \sim 10^{12}-10^{13}$G, with a dipolar $1/r^3$
falloff to higher altitudes.}
 above the star's surface.  Within the
so-called ``light cylinder'' -- the radius at which the
corotation speed becomes lightspeed -- the magnetic field geometry is
thought to be nearly that of a vacuum dipole, rotating with the star
and having its magnetic moment oriented at some oblique angle to the
star's rotation axis. Approaching the light cylinder, relativistic
effects ``sweep back'' the dipolar field lines, generating a toroidal
component which becomes an outward propagating EM wave past the light
cylinder (Deutsch 1955).  Most poloidal field lines which leave the star's
surface return to the pulsar well within the light cylinder, creating
the ``closed field zone''.  It is generally thought 
that plasma in this region, pulled from the star's surface by rotation-induced
electric fields, becomes charge separated in just the amount needed to
short out any parallel electric field.  This necessary charge density is
known as the ``Goldreich-Julian
charge density'',  $n_{\rm GJ} = \Omega_* \cdot \bold B / 2 \pi e c$,
from a seminal {\it ansatz} by Goldreich \& Julian (1969, ``GJ'').
  If this density is attained, it supports
 corotation of a force-free plasma magnetosphere in the
closed-field-line region. It is also generally agreed that magnetic
field lines or flux ropes which originate near the star's magnetic poles
cross the light cylinder before they return to the star, creating an
``open field line'' region.  The area on the star's surface traced by the
outermost open field lines is called the ``polar cap''. 

\begin{table}
  \begin{center}
\def~{\hphantom{0}}
  \begin{tabular}{ccccc}
      altitude  & $B$   &   $\nu_{\rm B}$ & $n_{\rm GJ}$ & $\nu_{\rm p}(n_{\rm GJ})$
\\[3pt]
   $R_* \simeq 10 $ km &   $4 \times 10^{12}$ G & $1 \times 10^{19}$ Hz
    & $8 \times 10^{12}$ cm$^{-3}$ & $ 3\times 10^{10}$ Hz
\\[2pt]
   $R_{\rm LC}/2 \simeq 80 R_*$ &  $ 8 \times 10^9$ G & $ 2 \times 10^{16}$ Hz
& $ 2 \times 10^7$ cm$^{-3}$ & $4\times 10^7$ Hz
\\[2pt]
   $R_{\rm LC} \simeq 160 R_*$ &  $ 1 \times 10^6$ G & $ 3 \times 10^{12}$ Hz 
& $ 2 \times 10^6$ cm$^{-3}$ & $1 \times 10^7$ Hz
\\
  \end{tabular}
\caption{Scaling numbers for the Crab pulsar, assuming the 
Goldreich-Julian (``GJ'', 
Sec. \ref{EmissionZonesHighandLow}) model holds.  $\nu_{\rm B}$ is
the lepton Larmor frequency;   $\nu_{\rm p}$ is the lepton plasma frequency.
Values are given at the star's surface and
at two higher altitudes more relevant to the radio and high-energy
emission.  $R_* \simeq 10$ km is the radius of the neutron star,
 which rotates at
 $\Omega_* \simeq 190$ rad/s (rotation period 33 msec).   The distance
$ R_{\rm LC} = c / \Omega_* \simeq 160 R_*$
 is the light cylinder radius. The magnetic field
is assumed to be dipolar.  The number 
density of charge  required to maintain corotation is
 $n_{\rm GJ} \simeq \Omega_* B / 2 \pi c e$. In many applications the 
number density of the plasma is thought to be enhanced relative to the GJ
value, by a factor $\lambda = n / n_{\rm GJ}$.
 }
  \label{table:CrabParameters}
  \end{center}
\end{table}
 
This model is hardly the final word on the subject.  The structure
of the magnetosphere is complex, dynamic and nonlinear.  It is far
from clear that it will ever attain the steady-state situation
envisaged by Goldreich \& Julian, because magnetic fields, currents and
charges feed back on each other, creating a very dynamic
system.  Observations
reveal that the radio emission is variable on both short and long
timescales; this also suggests that the emission regions are
unsteady and dynamic.  Nonetheless, the model provides 
valuable insights, and it has 
 become part of the ``standard pulsar picture'' over the years.
To illustrate the model, and for later reference for the Crab pulsar,
we store important scaling numbers in Table \ref{table:CrabParameters}.

\subsection{The traditional radio geometry:  lighthouse beams}
\label{LighthouseBeams}

The picture just described has
 been particularly successful in providing a context for understanding 
the pulsed radio emission by which most pulsars have been discovered.
Plasma/current flow out along the open field lines enables
 the buildup of charge-starved regions (so-called ``gaps'').  Such gaps
may fill the open field line region 
close to the star's surface ({\it polar gap};  e.g.,
Ruderman \& Sutherland 1975), or may sit along the 
edge of the open field line region ({\it slot gap};  Arons \&
Scharlemann 1979). 
Parallel electric fields ($E_{\parallel}$; component along $\bold B$)
 in these gaps accelerate outflowing charges to
very high Lorentz factors (typically $\gamma \sim 10^6 - 10^7$). 
Curvature radiation from these fast charges supports magnetic pair
creation\footnote{In magnetic fields close to the critical field,
(the field at which quantum effects become important, 
$B_c = m^2 c^3 / e \hbar \sim 4.4 \times 10^{13}$G) a single energetic
photon can create an electron-positron pair.  Radiation from the daughter
 leptons in turn creates more pairs, which create more photons, and so on
... leading to a pair creation cascade. }
 in the strong magnetic field close 
to the polar cap. The pair cascade enhances the plasma density, relative
to $n_{\rm GJ}$, by a factor $\lambda = n / n_{\rm GJ}$.  Models suggest $\lambda
\sim 10^2 - 10^4$, and a final pair streaming speed $\g_s \sim 10^2 - 10^3$
 (Hibschman \& Arons 2001, Arendt \& Eilek 2002).
Thus, result of the pair cascade  is a fast particle 
beam moving through a slower-moving pair plasma, both
still moving out along the open field lines.

If some part of the energy in the fast outflow through the open field line
region is converted
to radio emission -- which will be strongly forward beamed, approximately
along the star's magnetic axis -- an observer will see a radio pulse
every time the star's rotation sweeps this ``lighthouse beam'' past
his or her sightline.  
In this simple picture, most pulsars should show only 
one pulse per rotation period, which is indeed the case.
 We would need special geometry to see
more:  either the magnetic moment and the viewing angle are 
both at $90^{\circ}$ to the rotation axis (giving two pulses $180^{\circ}$ 
apart), or both are very close to the
rotation axis (giving a complex pulse distribution if the emitting region is
inhomogeneous). This model also predicts the position angle of 
 linear polarization  should rotate as the radio beam
crosses the observer's sightline (Radhakrishnan \&
Cooke 1969; assuming the 
polarization direction is rigidly tied to magnetic field geometry in the
emission zone). Because both
 predictions hold true for the majority of well-studied
pulsars, this geometry has  become part of the standard pulsar picture for
the past several decades. 

\subsection{Expand  this picture: the extended magnetosphere}
\label{ExtendedMagsphere}

We've learned much more in recent years. The happy
conjunction of abundant high-energy pulsar observations
with significant advances in numerical simulations (both MHD and PIC
methods) have greatly broadened our understanding of the high-altitude
magnetosphere and its transition to the pulsar wind (known as the ``extended''
magnetosphere; e.g., Kalapotharakos \etal 2012a).  Various analytic
attempts to understand the region around the light cylinder
generally proved less than successful over the years,
 but several groups have now  modelled this region numerically. 
Initial numerical work assumed pair creation was
sufficiently abundant that the force-free condition ($\bold E \cdot \bold
B= 0$) holds everywhere (e.g., Contopoulos \& Kalapotharakos 2010,
 Bai \& Spitkovsky 2010).  More recently, modelling
has been extended to include phenomenological resistivity in regions
where current sheets are expected (e.g., Kalapotharakos \etal 2012b,
 Li \etal 2012  ). 

While details differ between authors, the basic properties of the extended
magnetosphere seem agreed on, and are in good agreement
with the analytic models of Bogovalov (1999).
Closed field lines and charge-separated regions exist within light cylinder, 
as predicted by the GJ model.   Plasma flows out 
open field lines, crosses the light cylinder, developing  into an outflowing
pulsar wind. Past the light cylinder, the plasma flow --
which must remain sublight -- trails the
star's rotation, creating a spiral pattern for the toroidal magnetic
field.  The poloidal field becomes asymptotically radial and develops
a ``split monopole'' structure, changing sign 
across an undulating equatorial current sheet. 
This picture of the extended magnetosphere reveals more regions which may
accelerate particles to relativistic energies and have the potential to be
emission zones for high energy emission, radio emission, or both.

{\it Gaps} are regions where force-free MHD breaks down and unshielded
$E_{\parallel}$ can exist; a variety have been suggested.  In addition
to the low-altitude polar gaps described above, some authors have
suggested {\it slot gaps}, which extend the polar gap from the star's
surface out to the light cylinder along the edges of the open field
line region (Arons \& Scharlemann 1979, Muslimov \& Harding 2004; also
called {\it two-pole caustics}, see next section.)  Other authors have
suggested {\it outer gaps} and {\it annular gaps}: vacuum regions
allowing $E_{\parallel} \ne 0$, originating close to the so-called
neutral charge surface (partway out along the open field region, where
the charge density needed for corotation changes sign), and extending
by some distance both towards the light cylinder and into the polar
flux tube (e.g., Cheng \etal 1986, Romani \& Yadigaroglu 1995, Qiao
\etal 2004).

{\it Current sheets}  have also been suggested as possible sites of particle
energization and high-energy radiation. In models of the extended magnetosphere,
current sheets exist both 
within the light cylinder -- where currents flow along the
separatrices between open and closed field lines -- and external to
the light cylinder -- where an undulating current sheet, frozen into the wind,
 separates the two hemispheres of the split-monopole wind. It is often
argued that reconnection events across the current sheets energize the
plasma, possibly to relativistic energies, making them additional sources
of high-energy emission (e.g., Petri \& Kirk 2005,
 Bai \& Spitkovsky 2010, Cerutti \etal 2015).  Although we have not seen much discussion of
these current sheets as sites of radio emission, we think the topic
warrants further study.

\subsection{Caustics, sky maps and high-energy emission}
\label{CausticsAndFriends}

Rapid recent advances in high-energy pulsar observations --
particularly the many $\gamma$-ray light curves obtained by {\it
  Fermi} (Abdo \etal 2013) -- reveal that double-peaked high-energy
pulsar light curves are the rule rather than the exception.  This
shows that high-energy emission cannot come from low-altitudes,
 because there are far too many double-peaked high-energy light
curves to be consistent with the special geometry needed (both viewing
and inclination angles close to $90^{\circ}$)  if the emission
were from low-altitude polar gaps.  It is now generally accepted
that the two
bright peaks typical of high-energy light curves come from extended
high-altitude regions -- gaps or current sheets -- associated with but
not local to the star's two magnetic poles.

The detailed structure of these high-altitude emission regions is not
understood, and cannot easily be determined from
observations. Finite photon light travel times, and
the non-dipolar magnetic
 field geometry at high altitudes, cause outgoing photons
emitted parallel to $\bold B$ to accumulate in
a few preferred directions (``caustics''). 
 Simulations document a highly nonlinear
mapping between the spatial structure of magnetospheric emission
regions and the rotation phases and viewing angles at which emitted
radiation is caught by the observer. In particular, double-peaked light curves
do not require special geometry, but are seen from a wide range of viewing and
orientation angles  (e.g, Dyks \etal 2004,  Bai \& Spitkovsky 2010).

How does all this connect to radio emission?  In many pulsars, the radio and
high-energy light curves are out of phase, with the main radio pulse
leading  the brighter of the two high-energy pulses by
$0.2-0.3$ of a rotation period. This is generally ascribed to radio emission
coming from low-altitude polar gaps, while high-energy emission comes from high
altitudes.  Because the emission mechanisms are also 
different, one might think the radio and high-energy emission zones 
 have little to do with each other.

The situation is different, however, for an interesting subset of radio
 pulsars, including the Crab pulsar.  In these stars  the main radio and
high-energy peaks occur at the same rotation phases.  This suggests
that both the radio and high-energy emission come from the same
spatial regions --  high-altitude gaps or current sheets -- 
sitting somewhere in the upper reaches of the magnetosphere.  
Although the location and structure of the emission regions are not
yet understood, comparison of radio and high-energy profiles gives
some insight.  The two high-energy pulses of the Crab pulsar
are much broader than their radio counterparts, and emission
bridges between the peaks are seen at high energies but not in radio (e.g.,
Abdo \etal 2010). There is a modest correlation between 
bright radio pulses and enhanced optical pulses
 (Shearer \etal 2012). Putting this all together, we infer that  radio emission
sites in the Crab pulsar are related to  high-energy emission sites, but
are more localized. Furthermore, the radio emission is highly unsteady; 
 it varies on short and long timescales. We infer that the radio emission
sites are dynamic, inhomogeneous regions.

\section{How do pulsars shine in radio?}
\label{HowDoTheyShine}

Although the standard picture described in Sec. \ref{EmissionZonesHighandLow}
can explain the phenomenology
of most radio pulsars, details of the physics are not well addressed.  The
key question --- unanswered despite several decades of hard work -- is
 just how and where the radio emission is produced.  The
problem is made challenging by the high brightness
temperatures\footnote{The brightness temperature -- a quantity
easily measured by
  radiometers -- measures the intensity at frequency $\nu$ of a radio
  beam: $k T_{\rm B} = I_{\nu} c^2 / 2 \nu^2$.  Brightness temperatures
  higher than any physically reasonable temperature point to coherent
  emission.  } of the radio emission (often quoted as $10^{26}$K, but
can be as high as $10^{41}$K; Hankins \& Eilek 2007).  Such high
intensities are unlikely to come from any incoherent emission process.
Instead, a so-called ``coherent'' process is needed, such as maser 
amplification or coordinated motion of a group of charges.

Three things comprise the ``emission physics'' needed to make the intense
radio emission we see from pulsars.

\begin{itemize}

\item 
There must be a
 robust, long-lived site within the magnetosphere where conditions are right
to make the radio emission seen at regular, well-defined phases of the
star's rotation (``pulses'' or ``components of the mean profile'').
  Polar gaps in the lighthouse model
are one example;  localized regions within extended high-altitude gaps
or caustics are another example.

\item
There must be a source of available energy which can be tapped for radio 
emission; that source must come from dynamics within the  magnetosphere.  One
 example is relativistic plasma outflow in the open field line
region, driven by electric fields parallel to the magnetic
field.  Another example may be conversion of magnetic energy, for instance
reconnection events in current sheets.

\item
There must be a mechanism by which the available energy is converted
to coherent radio emission.  While it is generally agreed that 
such a mechanism must operate at the microphysical (plasma) scale,
 the details of
the emission mechanism are not yet understood. 

\end{itemize}

A wide range of models have been proposed, over the years, to meet
these conditions and explain pulsar radio emission. None of them has
yet emerged as the definitive answer, in part because it has been
very hard to compare the models to the data.  Most of the models
assume a beam of relativistic charges provides the available energy,
but other variants have also been suggested.  There are great
differences between the models in how that energy is converted to
coherent radio emission, as well as in the plasma conditions required
for a particular mechanism to work.   We discuss several of these
models in Appendix \ref{App:RadioEmissionModels}, and test some of them
against our data for the Crab pulsar in Secs. \ref{Nanoshot_models}
and \ref{HFIP}, below.

\section{Radio emission from the Crab pulsar}
\label{CrabInRadio}

The well-studied pulsar in the Crab Nebula is an important case study, 
because the high 
radio power of its occasional ``giant'' pulses makes it an ideal target for 
dedicated radio studies, such as our group has carried out over the years. 
We have studied the star over a wide frequency range, capturing
individual bright pulses with time resolution as low a fraction of a
nanosecond.  We 
have also continuously sampled the star's radio emission at many
frequencies, from which we determine both the star's mean radio
profile and the unsteady nature of its radio emission. In this paper we
draw on this body of work to characterize
the Crab pulsar's radio properties and consider what they reveal
about conditions in the pulsar's radio emission zones.  The reader
interested in further details of our work can refer to Moffett \&
Hankins (1996, 1999); Hankins \etal (2003); Hankins \& Eilek (2007);
Crossley \etal (2010); Hankins \etal (2015) and Hankins \etal (2016).

\subsection{Mean profile of the Crab pulsar}

\begin{figure}
  \centerline{\includegraphics[width=0.99\columnwidth]{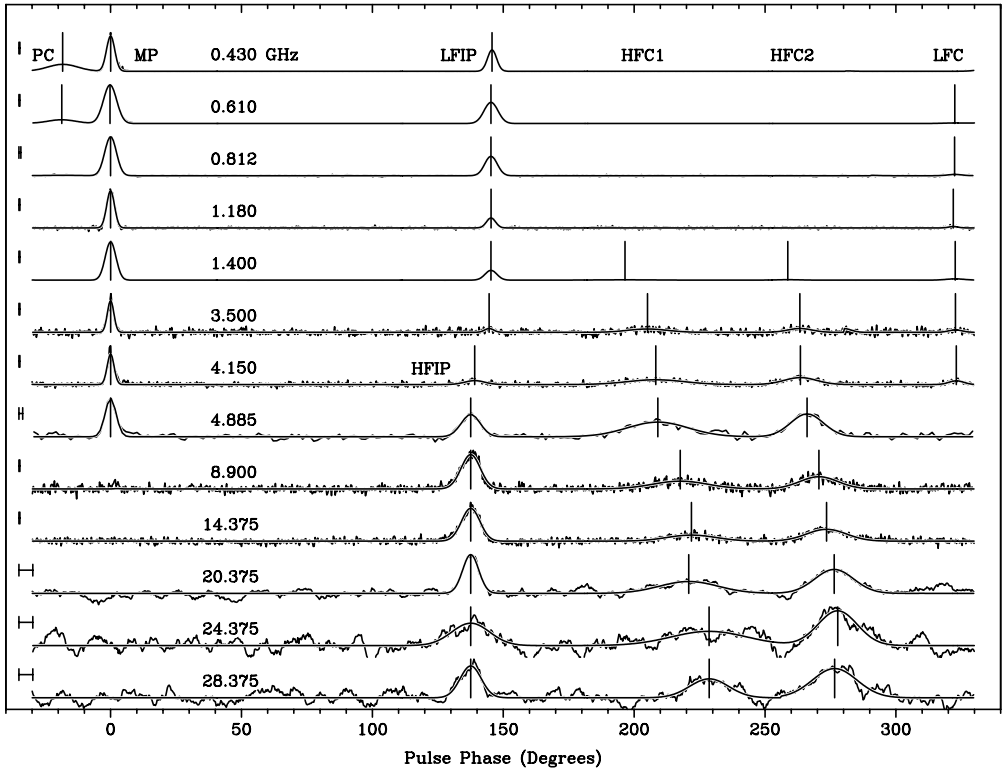}}
  \caption{Mean radio profiles for the Crab pulsar, at a number of frequencies,
with formal Gaussian fits overplotted.  The dominant
components discussed in this paper are described in the text and summarized with
acronyms in Table \ref{table:RadioComponents}.  The Main Pulse (MP) and the two
Interpulses (LFIP, HFIP) are coincident in phase  with the two peaks in 
the high-energy profile, which suggests a similar spatial origin for the radio
and high-energy emission.  The High-Frequency Components have not been clearly
detected in high energy profiles. Additional radio components seen at 
low frequencies, the Precursor (PC) and the Low-Frequency Component (LFC),
 may come from 
low-altitudes close to the star's polar cap. Nomenclature is
from Moffett \& Hankins (1996); figure from Hankins \etal (2015). }
\label{fig:MeanProfiles}
\end{figure}

The mean radio profile of the Crab pulsar is complex, with seven
distinct components identified so far.  The strength of each component
depends on frequency: radio observations above $\sim 5$ GHz
find a very different profile from that seen at lower radio
frequencies.  We show the radio evolution of the mean profile, with
components labelled, in Fig. \ref{fig:MeanProfiles}, and
summarize the components in Table \ref{table:RadioComponents}.

The Crab's mean 
profile at low radio frequencies -- between $\sim 100$ MHz and $\sim 5$
GHz -- is dominated by two peaks, the 
{\em Main Pulse}  and the {\em Low Frequency Interpulse}, separated by $\sim
145^{\circ}$ of rotation phase. Both are phase coincident with the 
two strong peaks in high-energy light curves (e.g., Abdo \etal 2010;
Zampieri \etal 2014).  However, both radio components 
are significantly narrower in phase than their high-energy counterparts,
and each radio pulse {\it lags} the corresponding high-energy pulse by
$\sim 2-3^{\circ}$ of phase.  Two 
additional weak components can be seen at low radio frequencies:
the {\em Precursor} leads the Main Pulse by $\sim 20^{\circ}$, and the
{\em  Low-Frequency Component} leads the Main Pulse by $\sim 37^{\circ}$.

The Crab's mean profile is quite different at high radio frequencies.
 It is dominated by an interpulse
and two new components  which are weak or
nonexistent below $\sim 5$ GHz.   The Main Pulse nearly disappears 
above $\sim 8$ GHz.  The Interpulse undergoes
a  discontinuous phase shift of $\sim 7^{\circ}$, and
also undergoes a  dramatic change in its spectral and
temporal properties of its single pulses.
  Because the differences between the
low-frequency and high-frequency Interpulses are so striking, we
identify them as two separate components. 
Despite the $7^{\circ}$ phase shift, the {\it High-Frequency Interpulse} is
also phase coincident with the second of the two broad peaks in the
high-energy light curve, but now this radio component {\em leads} the
high-energy component by $\sim 3^{\circ}$ of phase.  

Two new mean-profile components appear at high radio frequencies:  the
 {\it High-Frequency Components}. These first appear at a few GHz, and become
increasingly dominant going to higher frequencies.  By 28 GHz, the highest
frequency at which we have a mean profile, the two High-Frequency Components
are as strong as the High-Frequency 
Interpulse.  Unlike other components, the
 High-Frequency Components  drift in rotation
phase, appearing later in phase as one goes to higher frequencies.  There
is no clear sign of either High-Frequency Component in high-energy profiles, 
except for a weak possible detection in the $\gtw 10$ GeV profile of
Abdo \etal (2010).

The fact that we see seven radio components, distributed throughout
the full rotation period, immediately tells us  we are not seeing
low-altitude, polar cap emission.  If that were the case, we would see
at most two components, separated by $\sim
180^{\circ}$ of phase.  Furthermore, both the magnetic inclination
angle and viewing angle of the pulsar would have to be $\sim
90^{\circ}$ in order for two polar-cap components to be detected.
That requirement disagrees with our known viewing angle of $\sim
60^{\circ}$ relative to the star's rotation axis (determined from the
X-ray torus which surrounds the pulsar; Ng \& Romani 2004).  We thus
conclude -- consistent with arguments in
Sec. \ref{CausticsAndFriends} -- that the main radio pulses\footnote{
Possible exceptions are the Precursor and the Low-Frequency Component;
one of both of these may come from polar-cap emission.  See discussion
in Hankins \etal (2016).}  from the
Crab pulsar come from high altitude regions in the magnetosphere.

\begin{table}
  \begin{center}
\def~{\hphantom{0}}
  \begin{tabular}{cccc}
      Component & Acronym & Frequency Range & High-energy counterpart?
\\[4pt]
  Precursor & PC & 0.3--0.6 GHz & none found
\\[2pt]
  Low-Frequency Component & LFC & 0.6--4.2 GHz & none found
\\[2pt]
  Main Pulse & MP & 0.1--4.9 GHz & yes, P1
\\[2pt]
  Low-frequency Interpulse & LFIP & 0.1--3.5 GHz & yes, P2
\\[2pt] 
  High-frequency Interpulse & HFIP & 4.2--28. GHz & yes, P2
\\[2pt]
 High-frequency Components & HFC1,2 & 1.4--28. GHz & weak or non-existent
\\[2pt]
  \end{tabular}
\caption{Components of the mean radio profile of the Crab pulsar 
which we discuss
in this paper.  The frequency range describes the range over which 
components have been found in mean profiles from our group
(Moffett \& Hankins 1996, Hankins \etal
2015) or other work (e.g. Rankin \etal 1970).
  We have occasionally detected single Main Pulses and Low-Frequency
Interpulses  above the listed frequency
ranges, but they are too rare to contribute to  mean profiles.  Acronyms are
used in Fig. \ref{fig:MeanProfiles}. The two peaks seen in  
high-energy profiles
are P1 (``main pulse'') and P2 (``interpulse'');  they can be tracked
continuously from optical 
(S\l owikowska \etal 2009) to $\gamma$-rays (e.g. Abdo \etal 2010). }
  \label{table:RadioComponents}
 \end{center}
\end{table}

\subsection{Our single-pulse observations of the Crab pulsar}

Our main focus in this paper is what we have learned about individual
Main Pulses and Interpulses from our high-time resolution observations
of the Crab pulsar.  This work was
 carried out at between 1993 and 1999 at the Karl
G. Jansky Very Large Array, between 2002 and 2009 at the Arecibo
Observatory, and from  2009 and 2011 at the Robert C. Byrd Green Bank
Telescope.  Details of our single-pulse
observations and data-acquistion system
are given in Hankins \& Eilek (2007) and Hankins 
\etal (2015). Our net ``take'' was 1440 bright single pulses above $\sim 1$ GHz
with high enough signal-to-noise to be useful.  Pulses at lower
frequencies are significantly broadened by interstellar scintillation,
thus not useful for our analysis here.  We captured about 870 strong
Main Pulses,  540 strong Interpulses (510 of which were
High-Frequency Interpulses, the rest Low-Frequency Interpulses), and
30 High-Frequency Component pulses.  In this paper we focus on the
Main Pulse and the two Interpulses, for which our data allow useful
comparisons to the models.

Because our data-acquisition system is triggered only by the brightest pulses,
we have captured single pulses from 
the bright end of the pulse fluence distribution.  These are
sometimes called ``giant pulses'' in the literature.  However, there is
no evidence that these are any different from fainter, ``normal'' pulses
that make up the majority of the star's pulsed radio emission.  We therefore
assume the pulses we have captured are good representatives of the emission
physics for both high and low radio power.

Our single-pulse work shows that two 
very different types of radio emission physics exist in the magnetosphere
of the Crab pulsar. As we show in the following sections,
single Main Pulses and Low-Frequency Interpulses share the same
temporal and spectral characteristics, while single 
 High-Frequency Interpulses are very different.
This result was totally unexpected.  In Sec. \ref{EmissionZonesHighandLow} 
we argued that the Main Pulse and the two Interpulses should be
 associated  with the star's
two magnetic poles, probably from high-altitude caustics above each
pole.  However, because
no  pulsar model to date predicts any asymmetry between the 
north and south magnetic poles, there is no reason to expect the
radio emission physics from the two regions to be different. 
And yet that seems to be the case.

We emphasize that we have not seen any secular changes in the
properties of the Main Pulse or either Interpulse.  The characteristics 
we summarize here are robust and constant, over the nearly 20 years
we have observed this pulsar.
In the rest of this paper we present the detailed
characteristics of each type of radio component, and consider how well -- if
at all -- existing models can explain the observations.

\section{Main Pulse emission:  microbursts and nanoshots}
\label{MainPulseBursts}

Although the Main Pulse and Low-Frequency Interpulse components of the mean
profile are smooth and well-fit by Gaussians
(as in Fig. \ref{fig:MeanProfiles}),
 the story is very different when pulses are observed individually. 

\subsection{Explosive bursts of radio emission}
\label{Microbursts}

\begin{figure}
  \centerline{\includegraphics[width=0.5\columnwidth]
     {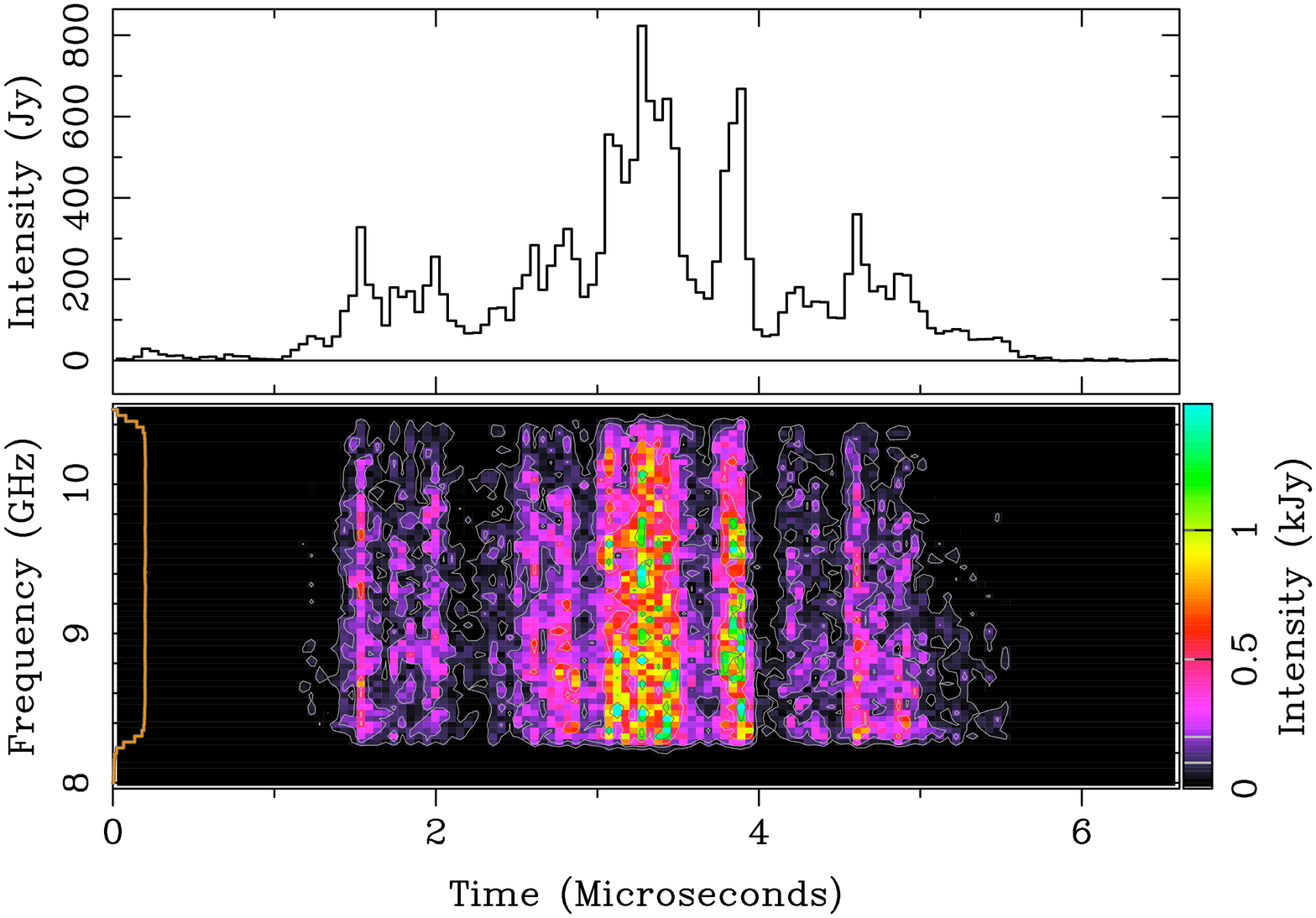}
        \includegraphics[width=0.5\columnwidth]
        {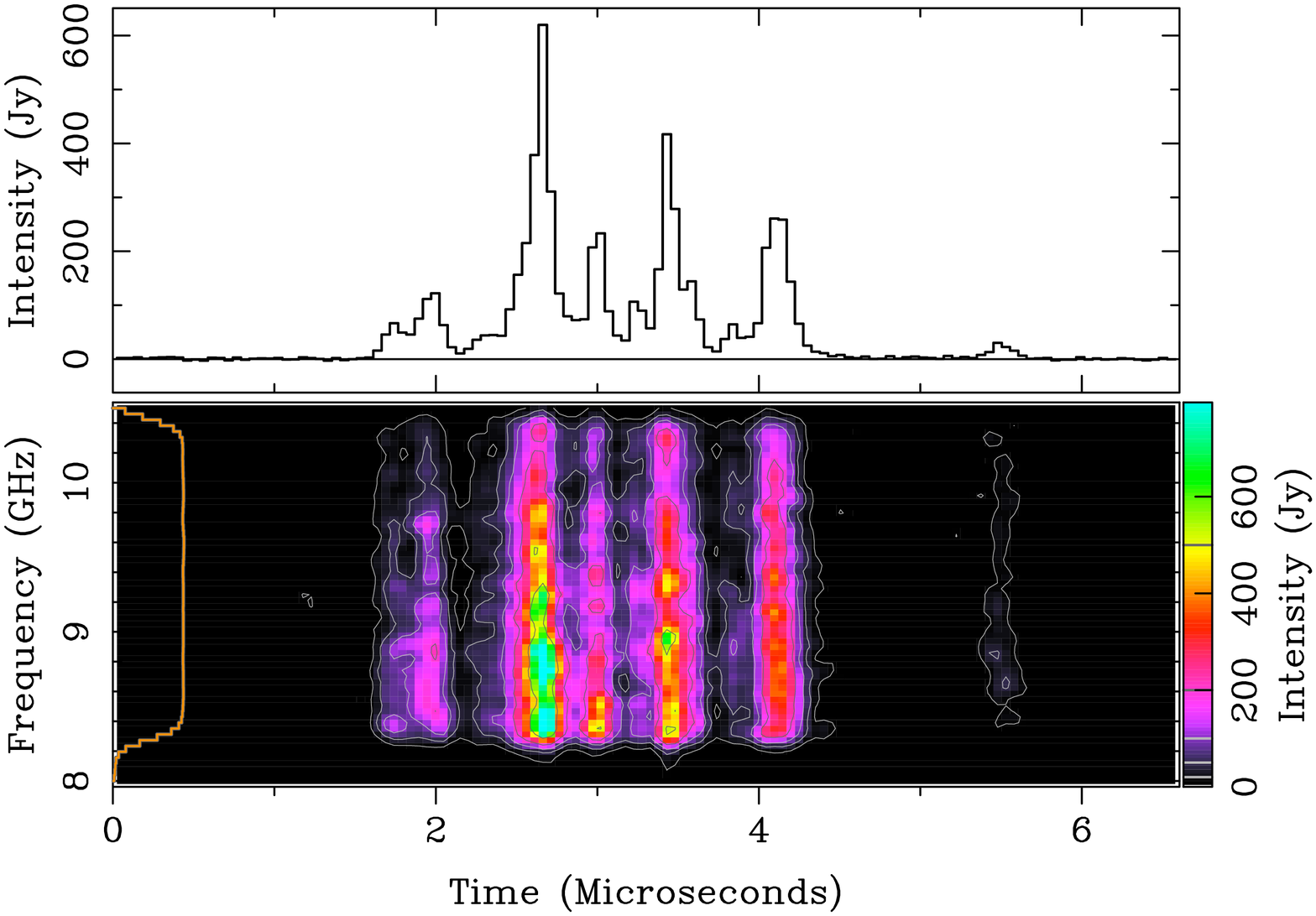}
}
  \caption{
Two typical examples of Main Pulses. Upper panels show intensity
integrated across our observing band.  The radio emission comes in
distinct bursts, each lasting less than a microsecond.  Lower panels
show the dynamic spectrum within our observing band; the orange line
on the left shows the equalized frequency response of the receiver.
The spectrum of each microburst is relatively broadband, spanning the
full observing bandwidth (but compare the pulse shown in
Fig. \ref{fig:MainPulse2}, where individual, narrow-band shots within
the pulse can be resolved).  Data were obtained with time resolution
equal to the inverse of the observing bandwidth. For display purposes,
both pulses have been smoothed to time resolution 51.2 ns and spectral
resolution 78 MHz. Both pulses were de-dispersed
with DM 56.73762 \pccm3 (see Sec. \ref{Dispersion} for definition of
DM). }
\label{fig:MainPulse1}
\end{figure}

The radio emission in a single Main Pulse or Low-Frequency Interpulse
comes in {\it microbursts}: abrupt
releases of coherent radio emission with a characteristic time of a
few microseconds.  Fig. \ref{fig:MainPulse1} shows two such examples.
These bursts can occur anywhere in the probability envelope defined by
the mean-profile component, which extends for several hundred microseconds
(Fig. \ref{fig:MeanProfiles}).  The burst strength is highly 
variable.  The strongest bursts are the most dramatic, but weak
bursts are more common. Furthermore, detectable bursts occur in random
groups that persist for a few minutes, between longer periods of radio
silence. 

The spectrum of a microburst is relatively broadband. As  Fig. 
\ref{fig:MainPulse1} shows, it  fills our observing bandwidth, which
was typically 2 or  4 
GHz for most of our observations, and extended to 8 GHz for one data
set. No lower frequency limit has been found for Main Pulse emission;
 that component can be seen in mean profiles well below 100 MHz 
(e.g., Rankin \etal 1970).  Main Pulses and  Low-Frequency 
Interpulses become faint and/or scarce above $\sim 10$ GHz.  They
disappear from mean profiles and we have captured very few
single pulses in these phase windows at higher frequencies.

\subsection{Sporadic energy release in the emission zone}
\label{SporadicBursts}

Microbursts show that the radio emission region is neither uniform nor
homogeneous.  It must be highly variable in space and time, producing the 
short-lived bursts of radio emission from isolated regions within the
emission zone.  We therefore need some process which sporadically
creates localized regions in which conditions are 
conducive to coherent radio emission.
We don't know the radio emission mechanism, but nearly all models
rely on a relativistic particle beam (as we
discuss in Secs. \ref{Nanoshot_models} and Appendix
\ref{App:RadioEmissionModels}).  The sporadic nature of
the radio bursts tells us the driving beams are not steady.  They must
dissipate after their energy is given up to the radio emission, and 
be continually regenerated throughout the emission region.

The driving beams must themselves be driven by electric fields.  In
highly magnetized regions such as the pulsar magnetosphere,
the driving $\bold E$  must be parallel to the local $\bold B$.  Thus,
$E_{\parallel}$ must itself be sporadic: reaching high strengths, 
creating the particle beam, then -- perhaps -- being shorted out.
This picture was originally suggested for the polar cap by  Ruderman 
\& Sutherland (1975), and has been explored numerically 
by Levinson \etal (2005) and  Timokhin \& Arons (2013). 
Their simulations verify that cyclic behavior can be a direct consequence of
low-altitude magnetic pair production, as
follows.  Say we start with a finite $E_{\parallel}$, driven by the star's
rotation, in a nearly empty  gap region.  It 
accelerates charges, which in turn create $\g$-rays by curvature
radiation.  The $\g$-ray photons decay into $e^+e^-$ pairs, which fill the
gap region and shield  $E_{\parallel}$.  Once the electric field 
is neutralized, particle acceleration ceases, the existing pairs stream out
of the region, and the original $E_{\parallel}$ is recovered as the gap
again becomes (nearly) empty.

The radio microbursts tell us that a similar process is needed in the
high-altitude radio emission regions of the Crab pulsar.  However, the
magnetic field is too low in those regions for magnetic pair 
 production to take place {\it in situ}. It may be 
that the particle beams which drive the radio emission
  have propagated from their natal regions
above the polar caps, but  it is not obvious that they would retain their
identity and still have the sporadic and inhomogeneous nature
required to explain the microbursts. Alternatively, 
two-photon pair creation may take place within the
 high-altitude gaps (e.g, Ho 1989, Cheng \etal 2000).   The seed
$\g$-rays are still thought to come from curvature radiation of charges
accelerated by unshielded $E_{\parallel}$ fields.  The target photons (likely
X-rays) may be thermal emission from the star's surface, or secondary
photons from a low-altitude pair cascade.  While this process has not
been studied in detail, it seems likely that an oscillatory $E_{\parallel}$
cycle can also be characteristic of high-altitude gap regions.

\subsection{Nanoshots}
\label{MP_nanoshots}

\begin{figure}
  \centerline{
\includegraphics[width=0.52\columnwidth]
  {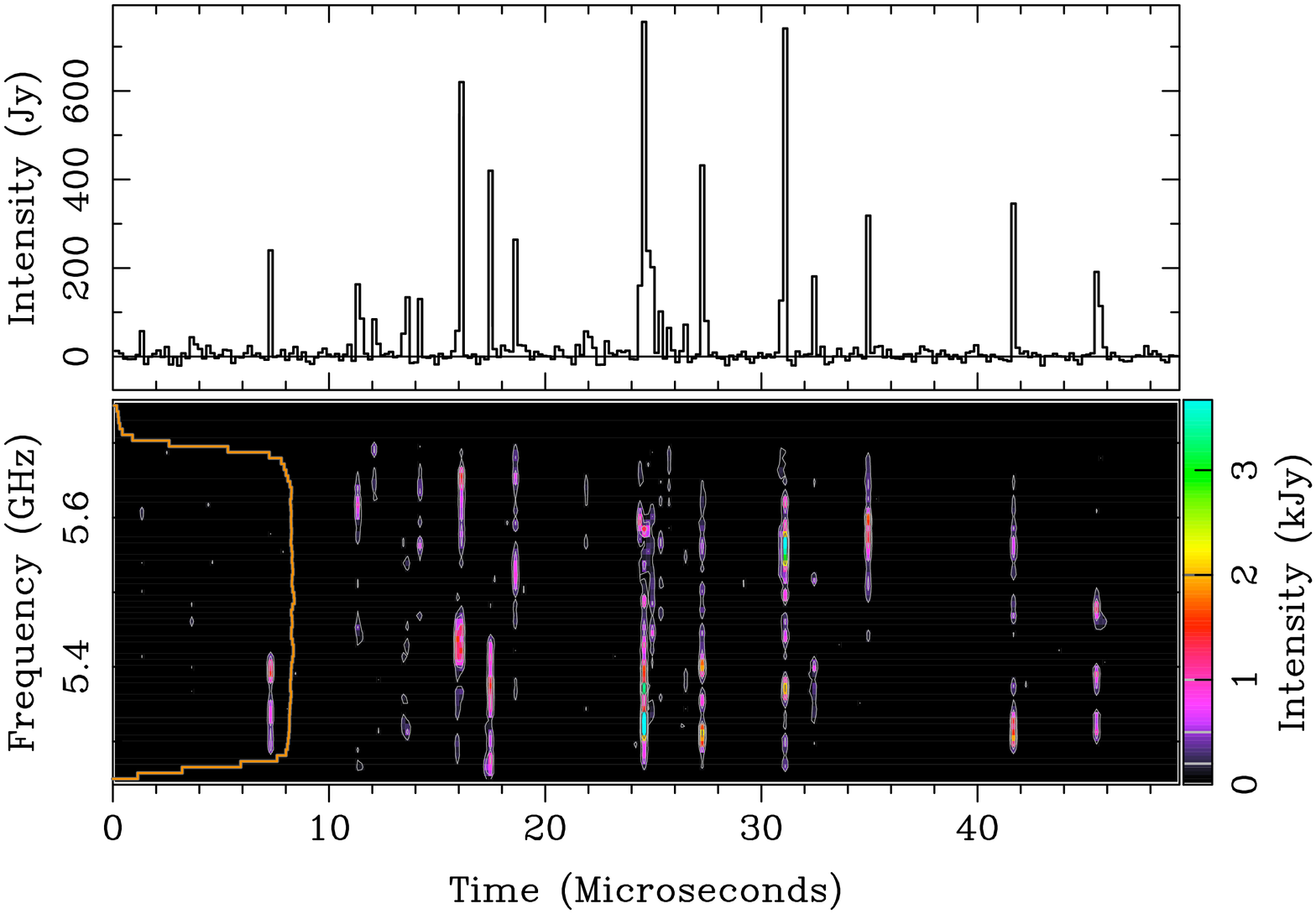}
\includegraphics[width=0.455\columnwidth]
    {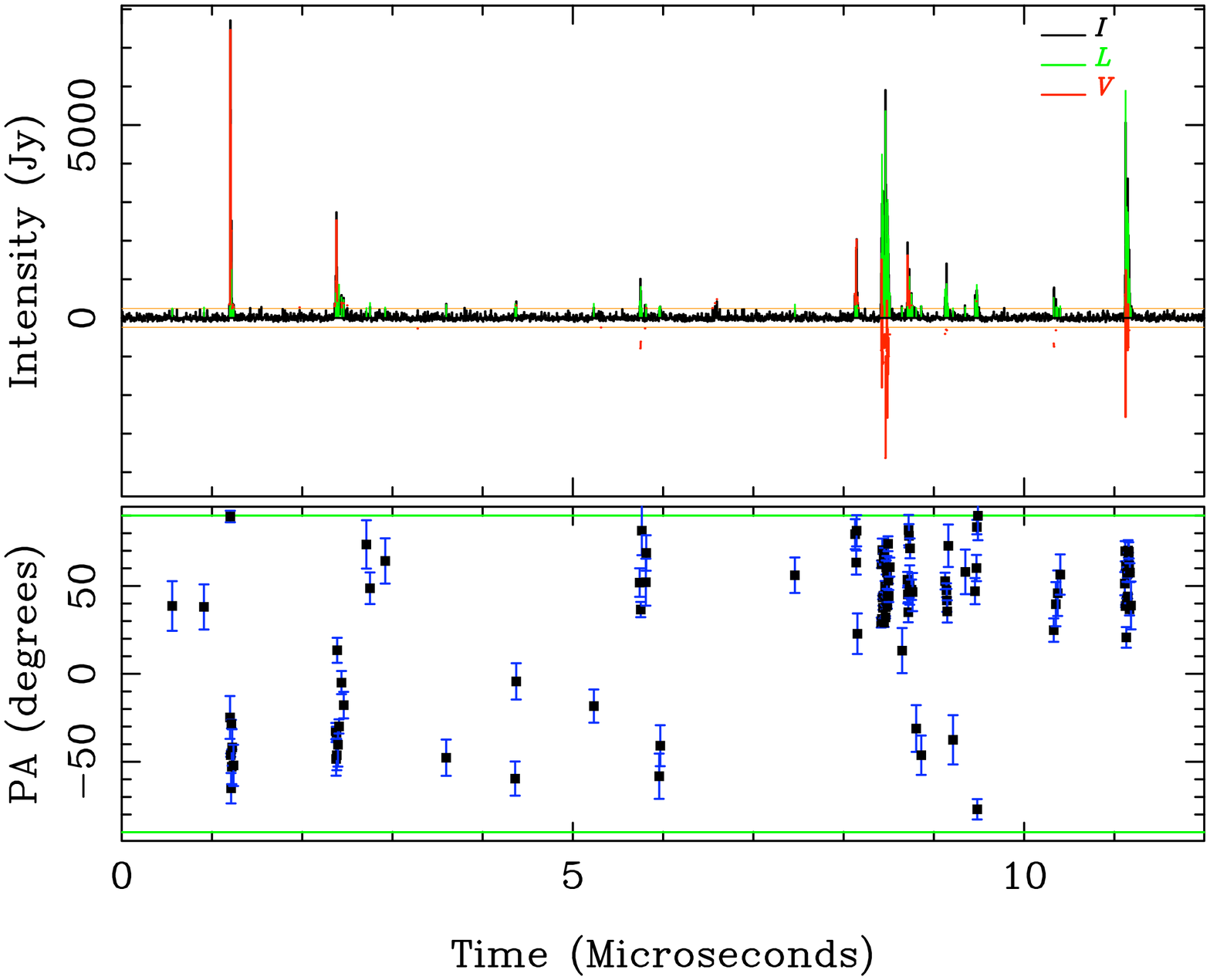}
}
  \caption{
An example of nanoshots in a Main Pulse.  Left figure shows the total
intensity and dynamic spectrum, de-dispersed with DM 56.76378 \pccm3;
layout is the same as in Fig. \ref{fig:MainPulse1}.  The radio emission in this
pulse is confined to a set of well-separated nanoshots.  Each nanoshot
lasts on the order of a nanosecond, and is relatively narrowband
(frequency spread $\delta \nu \sim 0.1 \nu$). Shown with time resolution 
51.2 ns and spectral resolution 39.1 MHz.   The right figure is a zoom
into the  nanoshots seen between 18 and 27 $\mu$s in the left figure,
now displayed at 4 ns time resolution.  At this resolution,
individual nanoshots within the clump at 24 $\mu$s are now resolved.
The 
different time smoothing in this view gives different peak fluxes for
each nanoshot. The upper panel shows the total intensity (I; black),
linearly polarized intensity (L; green) and circularly polarized
intensity (V; red).  The lower panel shows the position angle
of the linear polarization.  Only polarized flux above 4 times
the off-pulse noise (indicated by the orange lines in the top panel)
is shown.  Note the strong circular polarization of individual nanoshots
can be of either sign.
}
\label{fig:MainPulse2}
\end{figure}

Once in awhile a Main Pulse or a Low-Frequency Interpulse can be
resolved into very brief {\it nanoshots}.  We show  examples in Fig.
\ref{fig:MainPulse2};  others can be seen in Hankins \etal (2016). 
 Because we believe these
nanoshots provide a critical test of the radio emission mechanism, we
highlight their properties here. 

{\it Timescales.} 
Some of the nanoshots we have captured are
unresolved at our best time resolution (a fraction of a
nanosecond; the inverse of our observing bandwidth).  Others are
marginally resolved, lasting  on the order of a nanosecond.  The
example pulse we show in Fig. \ref{fig:MainPulse2} is substantially
smoothed, for clarity of presentation.  Nanoshots displayed at
higher time resolution can be found in Hankins \etal (2003) and  Hankins
\& Eilek (2007).

{\it Spectrum.}  
Each isolated nanoshot we have captured is relatively
narrowband, with fractional bandwidth $\delta \nu / \nu$ on the order
of 10\%.  Fig. \ref{fig:MainPulse2} shows that the center frequency
of the nanoshots is not fixed, but can be anywhere within our
observing band.  The product of the center frequency, $\nu_{\rm obs}$, and
duration, $\delta t$, of a nanoshot turns out to be
a useful diagnostic for
the models.  For each isolated nanoshot we have captured between 5 and
10 GHz this product $\nu_{\rm obs} \delta t \sim O(10)$.

{\it Polarization.}  
As the example in Fig. \ref{fig:MainPulse2}
shows, individual nanoshots tend to be elliptically polarized.  While
most of them
show some linear polarization, they are often dominated by
circular polarization (CP).  The sign of the CP can change from one
nanoshot to the next, apparently at random.

Although microbursts in 
most Main Pulses and  Low-Frequency Interpulses last longer and have
a broader dynamic spectrum than do the nanoshots,  we suspect that
microbursts are  incoherent superpositions of short-lived, narrow-band,
nanoshots. In other words, all microbursts are
 ``clumps of nanoshots''. This is supported by the observationally
motivated amplitude-modulated-noise model of pulsar emission
(Rickett 1975).  Because each such microburst is broadband
(Sec. \ref{Microbursts}), the center frequencies of the nanoshots within a
microburst must span a wide frequency range.

\section{Nanoshots as tests of radio emission models}
\label{Nanoshot_models}

It has been hard, historically, 
to compare radio emission models to the data.  The models address 
physical processes on plasma microscales, but most of 
the data can only address larger-scale, system-wide processes.
 We believe the nanoshots in the Main Pulse and Low-Frequency
Interpulse, revealed by our very high time resolution observations,  
provide a key clue to the underlying physics. 

In this section we use our data to confront
three plausible models from Appendix \ref{App:RadioEmissionModels}.
Because each model can be connected to  the spectrum,
duration and/or polarization of an individual nanoshot,
we can constrain  physical conditions which must hold in the radio
 emission region in order for that model to work.

For easy reference in the following discussion
 we refer back to the basic magnetospheric
parameters of the Crab pulsar in Table \ref{table:CrabParameters}.
We highlight the three models in Table \ref{table:NanoshotModels},
and summarize the plasma conditions required by each model in 
Table \ref{table:ModelParameters}.

\subsection{Strong plasma turbulence}
\label{SPT}

When electrostatic turbulence
is driven to large amplitude (which we call ``strong plasma turbulence'',
SPT), a modulational instability creates 
localized Langmuir-wave solitons.  These  solitons are themselves unstable, 
 generating electromagnetic waves which can escape the plasma, and potentially
be observed as nanoshots.  
We generally follow Weatherall (1997, 1998);
see also  Sec. \ref{App:SPT} of the Appendix.

{\it Spectrum.}  
Individual bursts are emitted around the plasma frequency
in the comoving plasma frame. If the plasma is moving out from the pulsar 
at a streaming speed $\g_{\rm s}$, we  observe nanoshots centered at
\be
\nu_{\rm obs}^{\rm SPT} \sim 2 \g_{\rm s}^{1/2} \nu_{\rm p}
\label{SPT_freq}
\ee
where $\nu_{\rm p}$ is the plasma frequency, and
everything is written in the observer's frame.  
Conditions required to put $\nu_{\rm obs}$ in the radio band
 are given in Table \ref{table:ModelParameters} and Sec. \ref{Nanoshot_summary}.
Simulations by Weatherall (1998) predict the nanoshots will be
 relatively narrowband, with $\delta \nu / \nu_{\rm obs} \sim \Lambda$,
  where $\Lambda$ measures the strength of the turbulence.
 Using $\Lambda \gtw 0.1$, from Weatherall (1997, 1998), 
this prediction agrees  with our observations of isolated nanoshots 
(Sec. \ref{MP_nanoshots}). 

{\it Timescales.} 
The dynamic timescale for soliton collapse, and the duration of the
consequent flare of radiation, is $\delta t \sim 1 / (2 \g_{\rm s}^{1/2} \Lambda
\nu_{\rm p}) $,  again written in the observer's frame. 
Using equation (\ref{SPT_freq}), 
this model predicts that radiation bursts observed at a few GHz  last
on the order of a nanosecond, and  have the product
$\nu_{\rm obs} \delta t \sim 1 /\Lambda$.  These predictions are also consistent
 with our observations of isolated nanoshots. 

{\it Polarization.}
 The collapsing soliton generates electromagnetic waves
whose polarization is set by local plasma conditions.  Weatherall (1998) 
worked in the context of a strongly magnetized plasma, in which charges
can only move along the magnetic 
field.  In this limit,  emergent radiation around
the plasma frequency is linearly polarized along the magnetic field. 
Because  nanoshots from the Crab pulsar can show strong circular
 polarization,  they cannot be explained by  Weatherall's model as it
 stands.   That model
would have to be extended, perhaps to a plasma in which the natural modes
are elliptically polarized.

\begin{table}
  \begin{center}
\def~{\hphantom{0}}
  \begin{tabular}{ccccc}
    Model  & Secs. & Spectrum & Timescales & Polarization 
\\[3pt]
   Strong plasma turbulence (SPT) & \ref{SPT}, \ref{App:SPT}
& consistent & consistent & needs work$^a$
\\[2pt]
   Free electron masers (FEM) & \ref{FEM}, \ref{App:FEM}
& plausible & consistent & needs work$^a$
\\[2pt]
   Cyclotron instability emission (CIE) & \ref{CIE}, \ref{App:CIE}
& challenged$^b$ & needs work$^c$ & consistent
\\
  \end{tabular}%}
\end{center}
\caption{Overview of nanoshot models discussed in the text.
$^a$Both SPT and FEM, as currently formulated, predict fully
linear polarization, which disagrees with the observations.
$^b$The densities needed for CIE to match observations are much higher
than magnetosphere theories predict. 
$^c$The current formulation of CIE cannot address
the short-lived nanoshots we observe.  
}
  \label{table:NanoshotModels}
\end{table}

\subsection{Free-electron-maser emission}
\label{FEM}

A relativistic electron beam, moving parallel to the magnetic field,
generates Langmuir turbulence as usual.  The interaction of the beam
particles with that turbulence leads to coherent bunching of the beam
charges, and consequent strong bursts of radiation.  This model is
similar to the SPT model, but here the {\it beam} itself is what
radiates; emission from the background plasma is not considered.
See also  Sec. \ref{App:FEM} of the Appendix.
  
{\it Spectrum.}
Bunched electron beams, moving at $\g_{\rm b}$,
 scatter on intense, localized electrostatic waves
of the plasma turbulence.  The scattered radiation can be treated as
 inverse Compton scattering of the Langmuir photons.  If the plasma
is at rest,
  the scattered photon frequency is\footnote{If plasma is itself streaming 
at $\g_{\rm s}$,  the boosted radiation is seen in the observer's frame as
$\nu_{\rm obs}^{\rm FEM} \simeq  \g_{\rm b}^2  \g_{\rm s}^{-3/2} \nu_{\rm p}$,
which adds another free parameter to the model.}
\be
\nu_{\rm obs}^{\rm FEM} \sim 2 \g_{\rm b}^2 \nu_{\rm p}
\label{FEM_freq}
\ee
(e.g. Benford 1992).
 Conditions required to put $\nu_{\rm obs}$ in the radio band
 are given in Table \ref{table:ModelParameters} and Sec. 
\ref{Nanoshot_summary}.

{\it Timescales.}
Short-lived radiation bursts arise naturally in a free-electron maser,
when the beam particles interact with electric fields in 
the Langmuir turbulence they have generated.  Both 
the coherent
charge bunching, and the consequent radiation bursts, are characterized
by the plasma timescale for the background plasma: 
 $\delta t \sim 1 / \nu_{\rm p}$
 (e.g., Schopper \etal 2003).  Thus, this model predicts
the product $\nu_{\rm obs} \delta t \sim \g_{\rm b}^2$.
 Clearly the beam cannot be 
too fast;  $\g_{\rm b}^2 \sim O(10)$ is needed to match the frequency-duration 
product of our observed nanoshots.

{\it Polarization.}
Both the oscillating electric field in the induced Langmuir turbulence,
 and the motion of
beam charges in that field,  are parallel to the local magnetic field.
It follows that radiation bursts are linearly polarized, also
parallel to $\bold B$  (e.g., Windsor \& Kellogg 1974). This model would also
have to be extended to accommodate circular polarization.  
Perhaps the
bunched beam charges have finite and synchronized pitch angles.
This can work in the lab; 
Benford \& Tzach (2000) reported coherent synchrotron emission from bunched 
electrons in an incoming rotating beam.  However, it is not clear that such a
situation would arise naturally in a pulsar magnetosphere.

\subsection{Cyclotron instability emission}
\label{CIE}

A relativistic particle beam moving into a magnetized plasma generates
transverse waves when it couples to the plasma through the anomalous
cyclotron resonance.  Because the waves can escape the plasma without mode
conversion, this is  a direct emission process. 
This theory has been applied to pulsars by Kazbegi
\etal (1991, ``K91'') and  Lyutikov \etal (1999, ``L99'', and references
therein).  To compare this model to Main Pulse nanoshots,
we follow the specific problem setup assumed by those authors: 
 one-dimensional motion of a cold pair plasma, streaming outward
from the pulsar at $\g_{\rm s}$, as well as a faster ``primary beam'' 
moving through that  plasma at $\g_{\rm b}$.  
See also  Sec. \ref{App:CIE} of the Appendix.

{\it Spectrum.} Emission proceeds 
through the first harmonic of the anomalous cyclotron resonance.
The escaping radiation is around the resonant frequency: 
\be 
\nu_{\rm obs}^{\rm CIE}   = \nu_{\rm B} 
{ 4 \nu_{\rm B}^2 \over \nu_{\rm p}^2} { \g_{\rm s}^3 \over \g_{\rm res}}
\label{CycResFreq}
\ee
where $\nu_{\rm B}$ is the electron cyclotron frequency and 
$\g_{\rm res}$ is the Lorentz factor of the resonant beam particles
($\g_{\rm s} < \g_{\rm res} < \g_{\rm b}$). 
All terms are written in the observer's frame, and the solution describes
waves propagating nearly along the magnetic field.  Because this is a
resonant process, operating at the lowest harmonic, we guess the emission
is relatively narrow-band. 
 Conditions required to put $\nu_{\rm obs}$ in the radio band
 are given in Table \ref{table:ModelParameters} and Sec. 
\ref{Nanoshot_summary}.

{\it Timescales.}
Because this model has not been carried past the 
linear instability calculation, it cannot address short-lived 
nanoshots.  One would
expect the instability to generate resonant plasma waves, some remaining in
the plasma to generate pitch-angle scattering,  others potentially
escaping to be seen as radio emission. 
 Lyutikov considered possible
saturation mechanisms  (e.g. L99 \& references therein), but did not discuss
self-generated coherent particle bunching, which  would be needed
for bright radio nanoshots.  

{\it Polarization.}
If the distribution functions of the electrons and positrons in a pair
plasma are different, the normal modes of the plasma become circularly
polarized for propagation close to the magnetic
 field (Allen \& Melrose 1982, K91).  
The CIE models of K91/L99 take advantage of this fact by assuming relative 
streaming between the two species.  If the electrons are moving faster
than the positrons, their model predicts
 left-handed CP;  if the positrons are moving 
faster, one gets right-handed CP.

\subsection{Nanoshots:  implications for the emission region}
\label{Nanoshot_summary}

In this section we have compared three beam-driven radio emission models
  to our nanoshot data.  Each model has it own requirements on the plasma
density in the emission region, which we summarize here and in Table 
\ref{table:ModelParameters}.  We also point out that 
 each model requires further development
before it can match all of the observations.

\subsubsection{SPT and FEM models}

Both of these models involve
plasma  turbulence driven by a relativistic particle beam, presumably via
a two-stream instability. The SPT model emphasizes
radio emission from collapsing solitons when the turbulence has become
strong.  The FEM model emphasizes radio emission from coherent charge 
bunching caused by the beam 
responding to the turbulence it has created.   

In both models  the emission frequency
is related to the plasma frequency in the background plasma: one
boosted by $\g_{\rm s}$, the other by $\g_{\rm b}^4$.  In practice, of
course, we expect both phenomena to contribute to the 
nanoshots.  The relative importance of the two effects depends on the
relative densities of the driving beam and the background plasma, neither of
which we attempt to estimate.

To radiate between $100$ MHz and  $10$ GHz, 
the emitting plasma density and streaming or beam speeds must satisfy 
\be
n \g_{\rm s} \sim (3 \times 10^{7}  -  3 \times 10^{11}) {\rm cm}^{-3} 
\quad {\rm or} \quad 
n \g_{\rm b}^4 \sim (3 \times 10^{7} -  3 \times 10^{11}) {\rm cm}^{-3}
\ee
for SPT or FEM, respectively.
  Recalling that $\g_{\rm b}^2 \sim O(10)$ is needed to
match the 
frequency-duration product of the nanoshots, and that $\g_{\rm s}$ may be
as low as $\sim 10^2$ in some models,  we see that these two constraints
are not dissimilar.

Comparing these constraints to the GJ density in the upper
magnetosphere (Table \ref{table:CrabParameters}), we see that the
density enhancement, $\lambda = n / n_{\rm GJ}$, must be in the range
given by 
$10^2 \ltw \lambda \g_{\rm s} \ltw 10^5$ and/or $10^2 \ltw \lambda \g_{\rm b}^4
\ltw 10^5$ in the emitting region.  For modest values of $\g_{\rm s}$ and
$\g_{\rm b}$, the upper part of the required density (or $\lambda$) range is
generally consistent with models of pair production close to the polar
cap (as in Sec. \ref{LighthouseBeams}).  Perhaps pair cascades operating
in high altitude gaps result in similar enhancements and streaming
speeds.

The lower end of the density range is problematic, however.
Previous authors (e.g. Kunzl \etal 1998, Melrose \& Gedalin 1999)
have argued that models based on relativistic plasma emission have difficulty
explaining radio emission from young pulsars.  If the 
radio emission comes from low altitudes -- over the polar cap -- the
GJ density there is  too high,
 even with $\lambda \sim 1$, to be compatible with low radio frequencies.  
This problem is mitigated if the radio emission
comes from higher altitudes, as it does in the Crab pulsar. Perhaps the
lowest radio frequencies come from high altitude regions where the plasma
density is not significantly enhanced over the GJ density.

\subsubsection{CIE model} 

This model is also beam-driven, but the emission proceeds through the
cyclotron resonance, which becomes unstable in the low magnetic
fields found at high altitudes (for the specific model of K91/L99).  Because
this model includes relative streaming between electrons and positrons, it
can explain the circular polarization we observe in the nanoshots. 

Stringent conditions, however are required if
this model is to explain the radio data.  Matching the resonant
frequency (equation \ref{CycResFreq}) to the radio band requires 
\be
 (n \g_{\rm res} / B^3 \g_{\rm s}^2) \sim (1 \times 10^{40} - 1 \times 10^{42})
~ {\rm cm}^{-3} {\rm G}^{-3}
\ee
Even taking the most optimistic parameter choices within the standard
model ($B \sim 10^6$ G, $\g_{\rm res} \sim 10^6$ and $\g_{\rm s}$ of order unity),
we still need $n \sim 10^{16}-10^{18}$ cm$^{-3}$ in order for this
model to produce radio emission.

Such high densities require huge
pair enhancements, $\lambda \sim 10^9 - 10^{11}$, much higher
than predicted by models of polar-cap pair
cascades (Sec. \ref{LighthouseBeams}). For comparison, 
models of the Crab Nebula  suggest the mean outflow density from the pulsar
exceeds the GJ value  by 5 or 6 orders of magnitude
(e.g. Bucciantini \etal 2010). While that range also exceeds
the $\lambda$ values predicted by pair cascade models, the
density enhancement required if the CIE model is to match observations is
higher still.  Perhaps some process, such as magnetic reconnection, can
dump large amounts of mass into high-altitude emission regions.  Perhaps
unusually low magnetic fields -- which would lower the necessary density
 range -- exist in such regions. 

\subsubsection{What's missing?}

None of the three models, as currently formulated, can address
 all of the data. The SPT and FEM models can explain the existence of
the nanoshots, but not their polarization.  The CIE model can explain
the circularly polarized emission, but cannot explain the existence of
the nanoshots. 

We suspect the way forward is a combination of the two approaches.  If the
electrons and positrons have different distributions -- for instance,
 different streaming speeds -- the fundamental
modes of the plasma are circularly polarized.  Alternating signs of CP
in different nanoshots require alternating dominance of one or the other
species in each nanoshot.  Perhaps that can exist in 
a highly turbulent pair cascade zone.  Nanosecond-long bursts of coherent radio
emission suggest that the emitting charges are coherently bunched.  Perhaps
that can come from the nonlinear evolution of one or both of these models,
operating in a plasma which can carry circular polarization.

\begin{table}
  \begin{center}
\def~{\hphantom{0}}
  \begin{tabular}{ccc}
     Model & Conditions & Secs.
\\[3pt]
   Strong plasma turbulence (SPT) 
& $ n \g_{\rm s} \sim (3 \times 10^{7}  -  3 \times 10^{11}) $
cm$^{-3}$ & \ref{SPT}, \ref{Nanoshot_summary}, \ref{App:SPT}
\\[2pt]
   Free electron masers (FEM) 
&  $ n \g_{\rm b}^4 \sim (3 \times 10^{7} -  3 \times 10^{11}) $
cm$^{-3}$ & \ref{FEM}, \ref{Nanoshot_summary}, \ref{App:FEM}
\\[2pt]
   Cyclotron instability emission (CIE) 
& $n  \g_{\rm res} 
\sim (1 \times  10^{22} - 1 \times  10^{24}) B_6^3 \g_{\rm s}^3 $ cm$^{-3}$
& \ref{CIE}, \ref{Nanoshot_summary}, \ref{App:CIE}
\\
  \end{tabular}%}
\end{center}
\caption{Plasma conditions required for radio emission models discussed in
Sec. \ref{Nanoshot_models}  to match
the spectrum of the nanoshots.  The density of the pair plasma
is $n$;  for comparison, Table \ref{table:CrabParameters} shows
the GJ density $n_{\rm GJ} \sim (2 \times 10^6 - 2 \times 10^7)$ cm$^{-3}$
in the
upper magnetosphere.  The Lorentz factor $\g_{\rm s}$ describes the streaming
speed of the pair plasma;  $\g_{\rm b}$ describes the speed of the particle
beam that drives FEM emission, which must satisfy $\g_{\rm b}^2 \sim O(10)$ to
match the observations;    $\g_{\rm res}$ describes energy of the beam particles
which participate in the cyclotron resonance.  The magnetic field is
scaled to $10^6$G, the smallest field likely within the standard picture
of the magnetosphere.}
\label{table:ModelParameters}
\end{table}

\section{The High-Frequency Interpulse:  a different picture}
\label{HFIP}

When we first studied individual High-Frequency Interpulses, in 2004 through
2006, we were astonished to find  they have very different  properties
from Main Pulses and Low-Frequency Interpulses. Since then we
 have continued to study them, over several years, going to other
 telescopes and higher radio frequencies.  The result is
unchanged:  High-Frequency Interpulses are very different from Main
Pulses and Low-Frequency Interpulses, in ways no current model
of pulsar radio emission can explain.

\subsection{Microbursts  but no nanoshots}

\begin{figure}
  \centerline{\includegraphics[width=0.5\columnwidth]
   {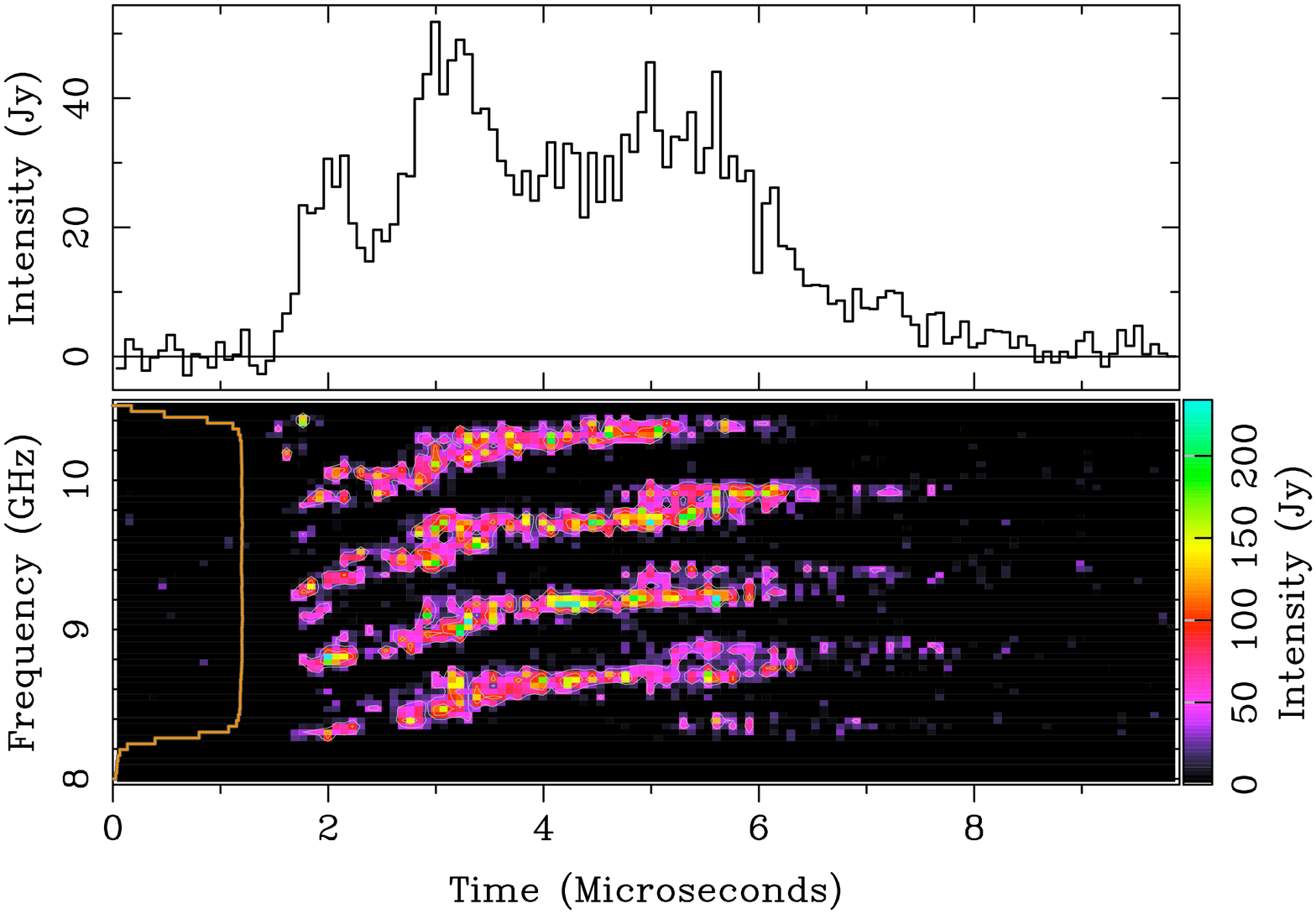}
\includegraphics[width=0.5\columnwidth]
     {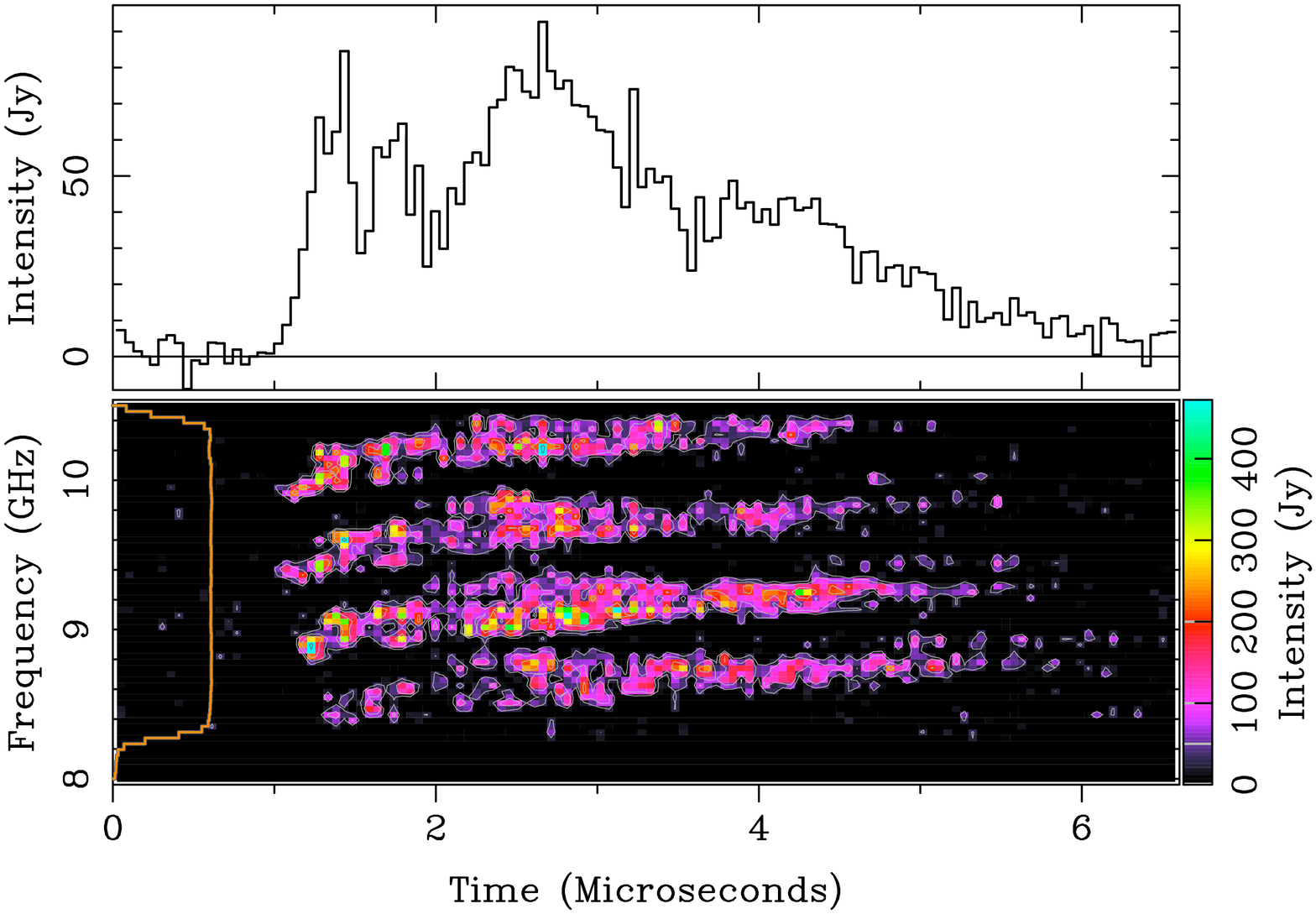}}
  \caption{
Two typical examples of High-Frequency Interpulses, shown with the same
layout as in Fig. \ref{fig:MainPulse1}.   Each pulse contains a small number of
bursts, each one lasting a few microseconds.  Comparison to
Fig. \ref{fig:MainPulse1} shows that bursts in this component are fewer, and
longer-lived, than in a typical Main Pulse.  The dynamic spectra show
the radio emission is concentrated in narrow {\it spectral emission
  bands}.  These  bands are very different from the Main Pulse spectra
(Figs. \ref{fig:MainPulse1} and \ref{fig:MainPulse2}), 
and were  not predicted by any model of pulsar radio emission.
  Each microburst in a High-Frequency Interpulse has its own
band sets: four sets can be seen in the pulse on the left, and
(probably) three can be found in the pulse on the right.
Pulse on left was de-dispersed with DM 56.73743 \pccm3 and 
displayed at 76.8 ns time resolution.  Pulse on right was de-dispersed with DM
56.75017 \pccm3 and displayed with time resolution 51.2 ns. Both pulses
displayed with 78.125 MHz spectral resolution.}
\label{fig:twoHFIPs}
\end{figure}

The temporal signature of High-Frequency Interpulses is similar to that
of Main Pulses and Low-Frequency Interpulses (Sec. \ref{MainPulseBursts}),
but with important differences.  Single High-Frequency Interpulses
 typically contain two or three
microbursts, each lasting a few  microseconds and overlapping in time. Fig.
\ref{fig:twoHFIPs} shows two examples.  As with Main Pulses and Low-Frequency
Interpulses, individual High-Frequency Interpulses are much shorter-lived
than the corresponding component in the pulsar's
 mean profile.  Single pulses can occur anywhere within
the mean-profile envelope; more than one pulse,
typically separated by a few hundred $\mu$s,  can occur in the same
rotation period. 

Also as with the Main Pulse and Low-Frequency Interpulse, the 
emission region for the High-Frequency Interpulse is 
dynamic.  Localized hot spots, with conditions favorable 
to radio emission, are created intermittently throughout the broad emission
region.  Each hot spot emits a burst of coherent radio emission,
then dies away, in only a few $\mu$s. Similar hot spots are 
regenerated again and again throughout the extended
emission region that is the source of the High-Frequency Interpulse.

There are, however, some differences between the time signature of  
High-Frequency Interpulses, and that of  Main Pulses and
the Low-Frequency Interpulses. 
 Each microburst in a High-Frequency Interpulse lasts about ten times
longer than that in a Main Pulse (compare Figs.
\ref{fig:MainPulse1} and \ref{fig:twoHFIPs}).  We have never recorded
a High-Frequency Interpulse
 with the well-separated, sub-$\mu$s bursts that characterize the
Main Pulses, nor have we ever captured nanoshots in a
High-Frequency Interpulse.   These details
suggest to us that  the energy
storage/release mechanism -- be it particle beams driven by $E_{\parallel}$,
or something completely different  -- is not the same in the two types
of emission regions.

\subsection{Spectral emission bands}

The unusual character of High-Frequency Interpulses is
clearest from their spectra. To quote Zheleznyakov \etal (2012), the
spectra revealed by our observations are ``completely unexpected
against a backdrop of forty-year-long studies of pulsar radio
emission''.  

Comparing the lower panels of Figs.
\ref{fig:MainPulse1} and \ref{fig:twoHFIPs} illustrates the
difference.  Where the burst spectrum of a Main Pulse or Low-Frequency
Interpulse is continuous across our observing band, the spectrum of a
 High-Frequency Interpulse
 is characterized by {\it spectral emission bands}.  Each
burst in a High-Frequency Interpulse contains its own band sets.  For
pulses with more than one burst and band set, the bands associated
with second or third  bursts tend to be at slightly higher frequencies
than those in the first burst (Fig. \ref{fig:twoHFIPs} shows good examples).

The spacing between adjacent bands is not constant, 
as one might expect for
simple harmonic emission. Instead, the band spacing increases with
frequency, while {\it keeping the fractional separation constant}.  
  Our data between 6 and 30 GHz are well fit by  $\Delta \nu \simeq 0.06 \nu$
(details in Hankins \& Eilek 2007, Hankins \etal 2016). 
Interestingly, our formal fit to the band spacing is consistent with
zero band spacing at zero frequency, even though we have seen no
High-Frequency Interpulses below 5 GHz.

We have never seen a High-Frequency Interpulse which did not have
emission bands across our full observing band.  Thus,
although we have never been able to observe
 the full 5-30 GHz bandwidth at one time,
we think it likely that each burst of a High-Frequency Interpulse
contains at least 30 emission bands, proportionally spaced from 5 GHz
upwards.

\subsection{Models for spectral emission bands}
\label{EmissBandModels}

The High-Frequency Interpulse of the Crab pulsar is the only place where
ordered
spectral emission bands  have been seen in a pulsar.  The bands cannot be
satisfactorily explained by any current model of pulsar radio
emission (including those listed in  Appendix \ref{App:RadioEmissionModels}). 

To search for new models, we note the resemblance between  emission
bands in the High-Frequency Interpulse and   ``zebra bands''
seen in  dynamic spectra of type IV solar flares.  Zebra band sets
can have from a few up to $\sim 30$ spectral bands;  they also have band spacing
that increases with frequency (e.g. Chernov \etal 2005).  Two classes of
models have been proposed for zebra bands:  geometric effects and
resonant plasma emission.  Perhaps of these models will help us  understand
the emission bands in the Crab pulsar.

\subsubsection{Geometric models}

The striking regularity of the bands suggests an interference or propagation
effect.  For instance, if some mechanism splits the emission beam coherently,
it may interfere with itself, creating emission bands.  Alternatively, perhaps
cavities form in the plasma and trap some of the incoming radiation,
imposing a discrete frequency structure on the escaping signal.  Both
possibilities have been suggested for solar zebra bands (e.g. Ledenev \etal
2001;  LaBelle \etal 2003).  The necessary plasma structures must be small,
on the order of a few to a few hundred centimeters.  

While we find this idea attractive, making it work is challenging.
One issue is the incoming radiation.  If
the bands are due to a geometrical effect, the incoming radiation must be
coherent over the full 5-30 GHz band.  The narrow-band shots which characterize
Main Pulse and Low-Frequency Interpulse emission do not work.  A quite
different emission mechanism is needed -- perhaps from a double layer.  Charges
accelerated in the potential drop of a double layer  emit broadband 
radiation (Kuijpers 1990).  If the double layer thickness is $\sim c / 
2 \pi \nu_{\rm p}$ (following Carlqvist 1982 for a 
 relativistic lepton double layer),
the radiation bandwidth may be sufficiently large. 

The more difficult challenge, however, is the basic geometry.  What 
long-lived plasma structures can create the necessary interference or 
wave trapping?  How can those structures be created or maintained so
consistently, within a dynamic magnetosphere, to give
 a steady $\Delta \nu / \nu$ which has not changed
over the seven years we have studied this star? We have not come up with
a good solution to this problem;  new ideas are needed.

\subsubsection{Double plasma resonance}

This model was also proposed to explain solar zebra bands 
(e.g.,  Kuijpers 1975, Zhelezniakov  \& Zlotnik 1975, 
 Winglee \& Dulk 1986).  Recent radio observations of solar flares 
(Chen \etal 2011) are in
good agreement with the model. If conditions are right (as detailed in 
Appendix \ref{App:DPR}), resonant  emission  occurs
when the plasma frequency is an integer multiple, $s$, of the electron
gyrofrequency:
\be
\nu_{\rm obs}^{\rm DPR}  \simeq \nu_{\rm p} \simeq s \nu_{\rm B}
\ee
 If the upper-hybrid turbulence excited by
the instability converts to electromagnetic modes which can escape
the plasma, we will see emission bands at discrete values of $\nu_{\rm obs}$.
In a non-uniform plasma the resonance for each integer $s$  occurs
at different spatial locations;  the band spacing depends on the specific
structure of the plasma density and magnetic field. 
Zheleznyakov \etal (2012)  suggested this mechanism can also account for
the emission bands in the High-Frequency Interpulse.  While we also find
this model promising, two problems are apparent.

One problem is the plasma setting.  The double plasma
resonance requires superthermal particles moving {\it across} a
weak magnetic field (with $\nu_{\rm B} \ll \nu_{\rm p}$).
This situation arises naturally in the magnetic bottles created by
solar coronal loops, from which only particles with low pitch angles can
escape.  However, such geometry is not part of our standard picture
for the pulsar magnetosphere, where particle beams moving {\it along}
the magnetic field are thought to be common (e.g., Secs.
\ref{HowDoTheyShine} or \ref{SporadicBursts}).  Perhaps 
magnetic traps can form in dynamic regions of the upper magnetosphere,
or associated with neutral current sheets (e.g. Zheleznyakov \etal
2012).

A second problem is the numbers.  To put the resonant frequencies 
in the radio band, the plasma density must be high and the
magnetic field extremely low. 
To get $\nu_{\rm obs}  \sim 5 - 30$ GHz, the plasma
must have $n \sim 10^{11} - 10^{13}$ cm$^{-3}$ and $B \sim 10^2 - 10^4$ G.  
Referring  to Table \ref{table:CrabParameters}, we see these parameter
ranges do not easily fit anywhere in the standard picture of the
magnetosphere. Higher
densities and much lower fields are needed.  Perhaps the necessary
magnetic loops or current sheets, if they do exist in the upper magnetosphere,
have the right parameters;  but no  models yet predict such structures. 
 Once again, new ideas are needed.

\section{Emission region of the High-Frequency Interpulse} 
\label{HFIP_environment}

The spectral emission bands are not the only unusual property of the
High-Frequency Interpulse. Unlike Main Pulses and Low-Frequency Interpulses,
 High-Frequency Interpulses are partially
dispersed {\it within the magnetosphere} before they leave the pulsar.  
In addition, their linear position angle shows no rotation across the entire
mean-profile component. Both properties give
us additional clues on the origin of High-Frequency Interpulses.

\begin{figure}
  \centerline{\includegraphics[width=0.95\columnwidth]
     {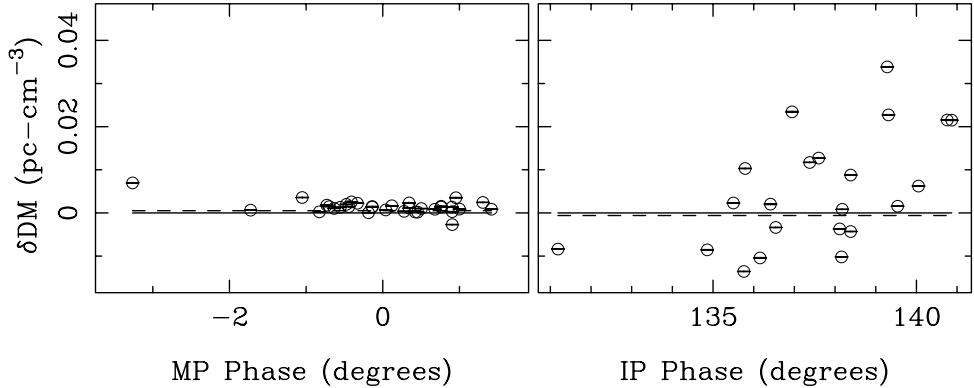}}
  \caption{ 
Excess dispersion measures, relative to Jodrell Bank monitoring, for
58 pulses captured within 80 minutes of each other during a typical
observing day.  Left panel shows $\delta  ({\rm DM})$ values for Main Pulses; 
 right panel shows $\delta ({\rm DM})$ for High-Frequency Interpulses.  
Observed at center frequency 6500 MHz, with bandwidth 1 GHz;
plotted against the rotation phase at which each pulse arrived
(compare the mean profiles in Fig. \ref{fig:MeanProfiles}).  
See Hankins \etal (2016)
for details of method.  The solid lines show the Jodrell Bank value 
(corresponding to $\delta ({\rm DM})  = 0$);
  the dotted lines are the mean $\delta ({\rm DM})$ for
each set of pulses. Dispersion of Main Pulses is
approximately consistent with the Jodrell Bank value, but
High-Frequency Interpulses have significant {\it intrinsic} dispersion
($\delta ({\rm DM}) \ne 0$), with large pulse-to-pulse scatter.}
\label{fig:DMscatter}
\end{figure}

\subsection{Intrinsic dispersion in the High-Frequency Interpulse} 
\label{Dispersion}

To quantify magnetospheric dispersion, we use 
the Dispersion Measure (``DM''), defined as follows.  The signal propagation 
time from a cosmic source depends on the group velocity in the
plasma through which it propagates: $ t_{\rm p}(\nu) \propto (d \om / dk)^{-1}$.
For cold, weakly magnetized plasma (such as the interstellar medium),
 the dispersion relation is well known, and we have 
\be
t_{\rm p}(\nu) \propto ({\rm DM}) ~\nu^{-2}  \quad {\rm where} \quad 
{\rm DM} = \int n dz
\label{DM_defn}
\ee
Thus, the difference in pulse arrival times at two frequencies 
 measures the column density between the signal source and us. 
This quantity is conventionally expressed as the DM, with units of \pccm3
(1 \pccm3 $ \simeq 3 \times 10^{18}$cm$^{-2}$). 

Nearly all of the dispersion for the Crab pulsar comes from the large plasma
column between us and the pulsar --  plasma in the Crab Nebula and the
intervening interstellar medium.  To focus on the small 
contribution from the pulsar itself, we express our results relative 
to the average DM which is tracked monthly by the Jodrell Bank Radio
Observatory.\footnote{Determined from relative arrival times
of bright components in mean profiles measured at 
610 and 1400 MHz;  http://www.jb.man.ac.uk/$\sim$pulsar/crab.html.} This
excess DM for an individual pulse, relative to the Jodrell Bank value,
 we call $\delta ({\rm DM})$. 

Fig. \ref{fig:DMscatter} shows $\delta ({\rm DM})$ values for Main 
Pulses and High-Frequency Interpulses on a typical observing day.  We find
 $\delta ({\rm DM})$ for Main Pulses is approximately
consistent with zero;  the Jodrell Bank values describe the Main Pulse well.
  However,  $\delta ({\rm DM})$ for
High-Frequency Interpulses differs significantly from zero and
 fluctuates strongly from pulse to pulse. It 
can vary from nearly zero to $\sim 0.04$ \pccm3, 
on a  timescale of a few minutes.  We note that observations of Main Pulses and
High-Frequency Interpulses  were interspersed with each other during 
a typical observing run.  No intervening
foreground ``screen'' could change  rapidly enough, and just in phase
with the pulsar's rotation, to exist only in front of the High-Frequency
Interpulse.   It is therefore clear that the excess DM of
the High-Frequency Interpulse comes from the pulsar itself, not an intervening
screen. 

We find no evidence that $\delta ({\rm DM})$ depends on observing
frequency between 5 and 30 GHz where we have captured
High-Frequency Interpulses.  This suggests the magnetospheric
dispersion law obeys $t_{\rm p}(\nu) \propto \nu^{-2}$ -- although it need
not be the cold plasma law assumed in equation (\ref {DM_defn}). We
note that a few $\delta ({\rm DM})$ values in Fig.
\ref{fig:DMscatter} are negative.  We do not interpret this as
inverted dispersion behavior ($ d t_{\rm p} / d \nu > 0$) in the pulsar.
Rather, it seems likely that the Jodrell Bank value, which is based on
mean profiles, contains a small contribution, the mean excess
dispersion, $\langle \delta ({\rm DM}) \rangle$, from the pulsar
itself.

Interpreting our result in terms of plasma conditions in the emission 
region is difficult, because we do not know the true dispersion law in
the region.  Magnetospheric dispersion relations have been well studied
only for low-altitude polar cap regions (e.g Arons \& Barnard 1986,
Melrose \& Gedalin 1999). While the effective DM has 
not been calculated for these laws, we suspect the more complex physical
situation in such strongly magnetized regions would cause more complex
$t_{\rm p}(\nu)$  behavior than the simple form in equation (\ref{DM_defn}).

As a possible example of dispersion in the upper magnetosphere,
consider dispersion in a relativistic, weakly magnetized plasma.  If
$\langle \g \rangle$ measures the mean internal energy of the plasma,
the DM over a path $L$ turns out to be 
$  \simeq n L / \langle \g \rangle$ (following, e.g. Shcherbakov 2008).
  Guessing $L \sim 0.1 R_{\rm LC}$ for a high-altitude emission zone, we need
$n / \langle \g \rangle \sim 4 \times 10^9$ cm$^{-3}$ to account for the
typical $\delta ({\rm DM}) \sim .02$ \pccm3 we see in the 
High-Frequency Interpulse.
Comparing this result to the high-altitude GJ densities 
(Table \ref{table:CrabParameters}), we find a density enhancement
$\lambda \sim (10^2 - 10^3)\langle \g \rangle$ would be needed.
This is generally consistent with predictions of pair cascade
models, as long as $\langle \g \rangle$ is not too large.  

The more interesting clue, however, is the $\delta ({\rm DM})$
variability. The plasma column around or above the emission site
 fluctuates by $n / \langle \g \rangle \sim 4 \times 10^9$cm$^{-3}$,
on timescales of no more than a few minutes.  This shows the 
 region creating the DM excess of the High-Frequency Interpulse 
is not some high-density region that happens to sit above the emission
zone.  It must be an intrinsic part of the dynamic emission zone. 

\begin{figure}
 \centerline{ 
 \includegraphics[width=0.495\columnwidth]
  {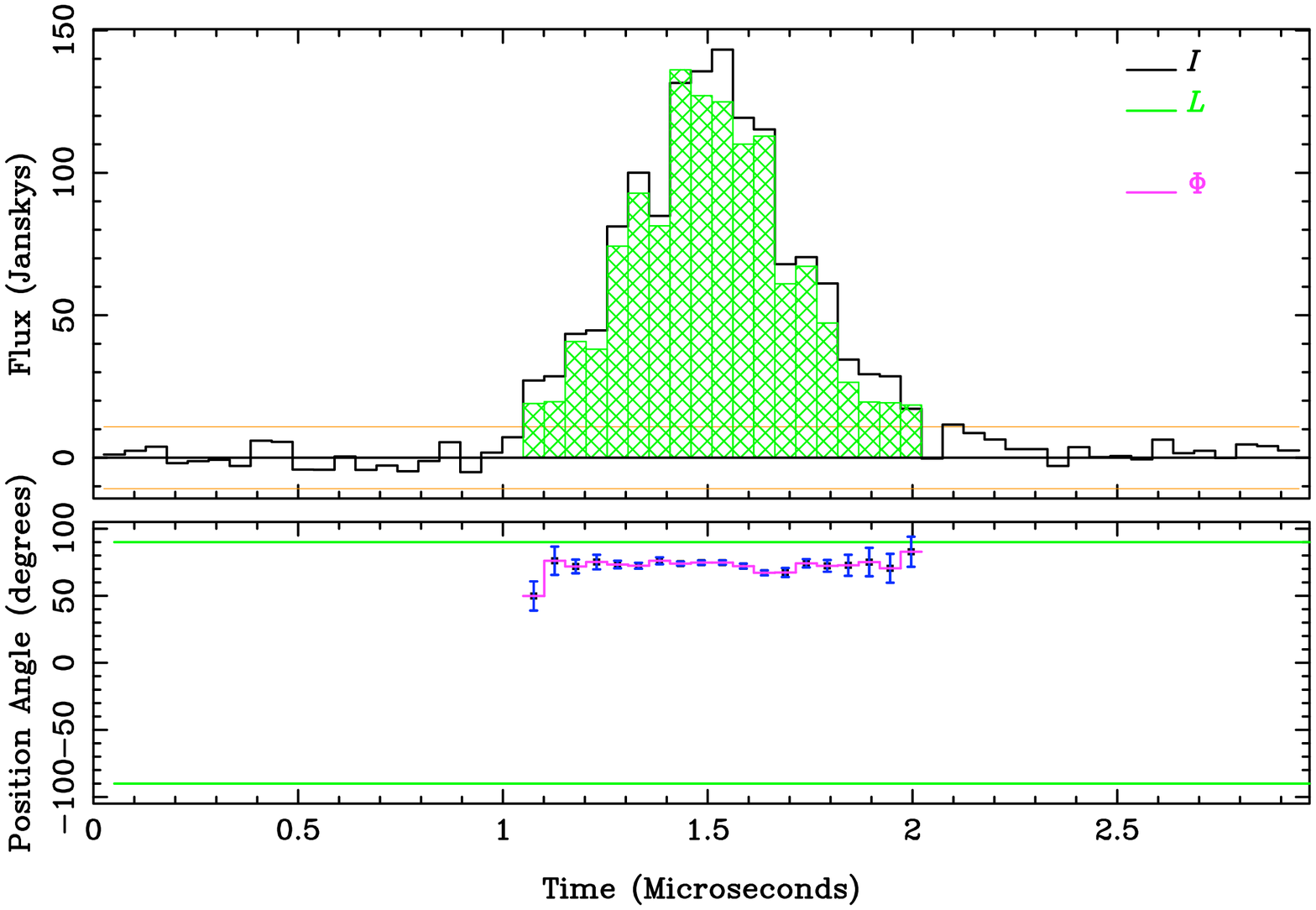}
\includegraphics[width=0.495\columnwidth]
  {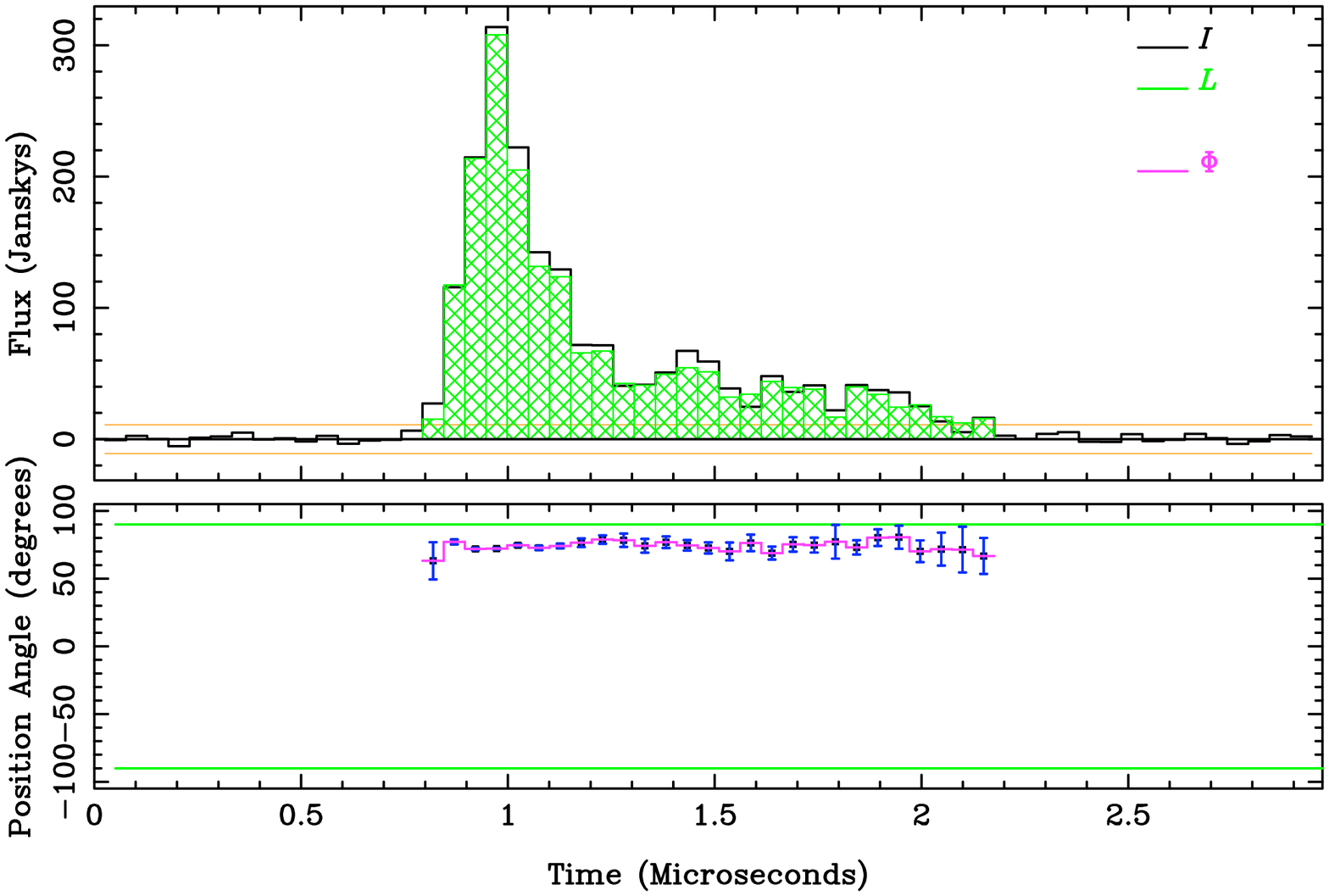}
}
  \caption{ Polarization  of two High-Frequency Interpulses, captured between  
17.75 and 22.75 GHz, within two minutes of each other on one observing day.
De-dispersed with DM 56.79476 \pccm3 and displayed with 51.2
ns time resolution. 
The pulse on the left arrived close to the leading edge of the 
High-Frequency Interpulse component in the mean profile 
(Fig. \ref{fig:MeanProfiles});
 the pulse on the right arrived close to the trailing edge. 
 For each pulse, the top panel shows the total
 flux (I; solid line) and linearly polarized flux (L; green filled area). 
The lower panel shows the position
angle of the linear polarization ($\phi$).  Polarization is
shown only  for points above three times the off-pulse noise (indicated by the
two orange lines in the top panel). 
}
\label{fig:HFIP_PA}
\end{figure}

\subsection{Linear polarization in the High-Frequency Interpulse}

High-Frequency Interpulses are linearly polarized.  Figure 
\ref{fig:HFIP_PA} shows two typical pulses, both of which are 
strongly polarized with nearly constant position angles. 
 The polarization provides two key diagnostics
of the emission region for this component.

Such strong polarization -- 90\% is typical -- constrains the extent
of the emission region.  We argued in Sec. \ref{CausticsAndFriends}
that the phase coincidence of the High-Frequency Interpulse and the
high-energy pulse P2 suggests the two features originate in similar
parts of the magnetosphere.  However, if the High-Frequency Interpulse
comes from a wide range of altitudes -- say throughout a slot gap or
an outer gap -- it would be depolarized by caustic effects (Dyks \etal
2004).  The strong polarization of the High-Frequency
Interpulse tells us caustic depolarization is not important for this
component.  Therefore, its emission zone must
occupy only a small fraction of the extended caustic region that is
probably the source of pulsed high-energy emission.

Figure \ref{fig:HFIP_PA} also shows that the polarized position angle
remains constant throughout a single pulse.  This  angle is independent
of the rotation phase at which the pulse appears. 
For example, the two pulses in Figure \ref{fig:HFIP_PA} arrived
at rotation phases separated by $6.5^{\circ}$ (which can be compared to
the $9.0^{\circ}$ width of the mean-profile component; Hankins \etal 2015), 
 yet both pulses show the same, steady polarization angle.

The position angle of pulsar radio emission 
reveals the direction of the magnetic field in the emission zone.
Radio emission is polarized 
 either parallel or perpendicular to the local field, depending on the
emission model.  The polarization angle
observed at earth therefore depends on the direction of the field as projected
on the sky plane. A change of angle through a pulse is 
 interpreted as a change of  magnetic field direction. 
The classic example is radiation from the low-altitude polar gap, which is a
region with diverging field lines.  As the star rotates 
past the line of sight, creating the radio pulse, the projected field
 direction  rotates across the pulse (e.g. Radhakrishnan \&
Cooke 1969).   

The lack of such position-angle rotation in the High-Frequency
Interpulse tells us the direction of the magnetic field is constant
throughout its emission region. The field does not diverge within the
region, as it does in the polar gap. Its direction does not change
with time, as might be expected in a turbulent region dominated by
plasma inertia.  The emission region for the High-Frequency Interpulse
must be inhomogeneous -- in order to account for the variable
dispersion -- but must also contain an ordered magnetic field.
Perhaps we are seeing a region filled with field-aligned density
fluctuations.

\section{Summary and lessons learned}
\label{SomethingAtTheEnd}

In this paper we have reviewed our high-time-resolution
observations of the Crab pulsar and compared them 
to various models for pulsar radio emission
physics. We offer hearty congratulations to the reader who has made
it this far. To summarize, our key results are as follows.

\begin{itemize}

\item  The bright radio 
pulses come from high altitudes, not from the polar caps.  Their emission
regions are probably 
co-spatial with the gaps or current sheets responsible for high-energy
emission. We 
use our data to probe physical conditions in the outer reaches of the
magnetosphere. 

\item Variability of the radio
signal on microsecond time scales, or longer,
 is generally consistent with the picture of cyclic pair cascades in an
unshielded gap. 
The details, however, remain unclear;  we suggest our radio data
 motivate new study of cascades in high-altitude emission regions.

\item Three emission models -- strong plasma turbulence,
free-electron masers, and cyclotron instability emission -- compare fairly
well to the nanoshots which characterize the Main Pulse and the Low-Frequency
Interpulse.  However, none of them
can explain all of the observations;  further work is needed.

\item The spectral emission bands which characterize
the High-Frequency Interpulse confound every radio emission model yet
proposed.  The model which comes closest is double-resonant emission, but
it requires extreme conditions which push the envelope of current
magnetosphere models.

\item  The variable dispersion and strong polarization of the
    High-Frequency Interpulse show the emission zone for that
    component is a localized, dynamic region that is probably threaded by a
    uniform magnetic field.

\end{itemize}
\smallskip

Our analysis makes it 
clear that we do not yet have all the answers about the 
pulsar's magnetosphere and how it makes coherent radio emission.  We have,
however, learned some useful lessons which should help us going forward.

$\bullet$ {\it The Crab pulsar has at least one radio
emission zone that is not predicted by current models.}

Strong radio pulses from the Crab pulsar come from localized zones
within extended high-energy emission regions -- probably gaps or
separatrices associated with each magnetic pole.  We infer this from
several facts.  The Main Pulse and Low-Frequency Interpulse
appear at the same rotation phases as the two strong components of the
high-energy mean profile, and the 
viewing angle of the star does not allow a low-altitude polar cap origin
for the latter.  Furthermore, these two components
have the same radio characteristics.  This suggests they come
from the same emission physics, which is consistent with their origin in
two similar regions connected to the two magnetic poles. 

The story is much less clear, however, for a third radio component.  The
High-Frequency Interpulse has very different characteristics from the
other two.  This suggests it comes from a region in which different
emission physics is operating, but we do not know where that is.
Because the High-Frequency and Low-Frequency Interpulses appear at similar
rotation phases, we might think they come from different parts of one
polar caustic.  However, the most likely  models suggest the
High-Frequency Interpulse arises in a dense, turbulent region
with uniform magnetic field direction. It is not clear where such a region
exists within our current picture of the  high-altitude magnetosphere.

$\bullet$ {\it High-altitude emission zones are unsteady places.
Their dynamical state is reflected in the radio emission.}

Radio emission from the Crab pulsar is not steady.  It comes in
short-lived bursts which can occur anywhere within the probability
envelope defined by the mean profile.
 We conclude from this that the mechanism which drives the
coherent radio emission exists only sporadically throughout the
emission zone.  Whatever drives the driver must also be sporadic.  The
energy source builds up, is  depleted to radio emission, 
and is soon regenerated to repeat the cycle. 

If the driver is a relativistic particle beam -- as is likely for the
Main Pulse and Low-Frequency Interpulse -- the duty cycle reflects
oscillatory behavior of parallel electric fields which drive the beams.
This picture is generally consistent with the  behavior expected
for unsteady pair cascades in gap regions.  If the Main Pulse and 
Low-Frequency Interpulse come from high-altitude caustic regions
associated with the star's two magnetic poles,  an unsteady
``sparking'' cycle must occur there, just as has been predicted for the
low-altitude polar cap.

The driver is less certain for the High-Frequency Interpulse, because
its radio emission mechanism is less clear.  Perhaps
unsteady reconnection is taking place in the mysterious region which
gives rise to this component.

$\bullet$  {\it The radio emission regions are either denser, or more tenuous,
than predicted by standard pair cascade
models, but we  do not know which.} 

Radio bursts in Main Pulses and Low-Frequency Interpulses
are clumps of nanoshots: short-lived, narrow-band, polarized
 flashes of radio emission. No models yet proposed 
can fully explain the nanoshots, but some come
close.  One class of models involves strong plasma turbulence, 
either directly (via soliton collapse) or indirectly (via the
bunched response of the driving beam).  These models require fairly
low densities within the emission zones -- comparable to, or somewhat
above, the GJ density.  An alternative model invokes
emission from a cyclotron instability.  This model requires substantially
higher densities in the emission zones, significantly above the GJ
density for the region and also well above densities predicted by
current  pair cascade models.  None of these models are the
final word on the nanoshots, but all of them appear to disagree with
predictions of the pair cascade models. 

Two other arguments also suggest high densities. 
While the emission physics behind the
High-Frequency Interpulse is not well understood, the promising 
double-resonant model  of the spectral bands 
 requires high densities in the  relevant emission
zone.  In addition,  some models of the Crab's pulsar wind nebula
 require high  densities in the  wind leaving the pulsar.
Neither of these models is likely to be the final word on
the High-Frequency Interpulse or the nebular wind, but both suggest
densities higher than pair cascade models can produce.  We conclude
that the pair cascade -- the only mechanism yet proposed to increase 
magnetospheric density over the fiducial GJ value -- is not yet understood.

$\bullet$  {\it  Radio emission can be a powerful probe
of magnetospheric conditions, but the models need more work.} 

We have discussed the radio emission 
models we find the most promising at present, and the limitations
of each one.  Models based on
plasma turbulent emission have been extended to nonlinear regimes, and thus
can address short-lived nanoshots.  However, they are 
restricted to electrostatic turbulence in strongly magnetized regions,
so they cannot account for circular polarization in the nanoshots. 
Models based on cyclotron emission have been constructed to admit circular
polarization.  However, they have not been developed past the linear
regime, so they 
cannot address the behavior of the nanoshots.  Models based on solar
zebra bands are a promising approach to the spectral emission bands, but they
require unusual physical situations which
are not part of our standard picture of the magnetosphere.

Our opinion is that these models -- which we have
highlighted -- can benefit by continued development, and help us truly
understand conditions in the high-altitude radio emission zones. 
These are not the only options, however.
Many other models, including those listed in Appendix 
\ref{App:RadioEmissionModels}, have promise but have
not been developed to the point where
they can be compared to data in any quantitative way.  Perhaps one of them
will turn out, in the end, to be a better answer for radio emission from
the Crab pulsar.

\bigskip

We are grateful to Gregory Benford, Maxim Lyutikov and Don Melrose for 
entertaining discussions and thoughtful reviews
that have helped our understanding and
appreciation of pulsar emission physics.  We particularly thank Glenn Jones
and many
of our students, including David Moffett, Jeff Kern, Jared Crossley,
Tracey Delaney, Eric Plum, Joe Dickerson, and James Sheckard for their
critical help at the telescopes during our observing sessions.  The
Arecibo Observatory is operated by SRI International under a
cooperative agreement with the National Science Foundation,
and in alliance with Ana G. M\'endez-Universidad
Metropolitana and the Universities Space Research Association.  The
Very Large Array and the Green Bank Telescope are run by the
National Radio Astronomy Observatory, a facility of the National
Science Foundation, operated under cooperative agreement by Associated
Universities, Inc.

\appendix

\section{Models proposed  for  pulsar radio emission}
\label{App:RadioEmissionModels}

A wide range of models have been proposed over the years to explain
the mechanism by which pulsars make coherent radio emission. The
models vary in both the plasma-scale processes
 responsible for the radiation and the magnetosphere conditions
under which they can operate. 

We do not attempt to give a full review of all possible radio
emission models.  The reader who has the strength can refer to the excellent
reviews by Melrose (such as Melrose 1995) for a more
complete discussion.   Instead, we present overviews of several models which
are commonly proposed and/or of particular interest for the Crab pulsar.

\subsection{Strong plasma turbulence}
\label{App:SPT}

This mechanism is known to operate in solar radio bursts, but must be revisited
for the high magnetic fields and relativistic pair plasmas relevant to pulsars.
An instability -- most likely two-stream, driven by an energetic
particle beam -- drives Langmuir turbulence. That
turbulence cannot escape the plasma directly; the Langmuir waves
 must undergo some nonlinear process
which converts part or all of their energy into escaping radiation. 

One possibility for this mode conversion involves modulational
instabilities in strong Langmuir turbulence that drive the growth
(``collapse'') of localized knots of strong electric field, generally
called solitons.  Nonlinear wave-wave coupling within the solitons
generates bursts of electromagnetic radiation -- ``nanoshots'' -- which can 
escape the plasma (e.g. Robinson 1997). 
  Weatherall (1997, 1998) has studied this process
in the pulsar context.  

 The duration and spectrum of the nanoshots provide key comparisons to
 the data.  In Weatherall's simulations, the rate of collapse of the
 electrostatic wave packet and subsequent conversion to
 electromagnetic modes is on the order of $\sim \Lambda \nu_{\rm p}$, where
$\nu_{\rm p}$ is the plasma frequency, and 
 $\Lambda = E_{\rm t}^2 / 8 \pi n m \bar {v_{\parallel}}^2$ measures the
 strength of the turbulence ($E_{\rm t}$ is the turbulent wave electric
 field, and  $\bar{v_{\parallel}}$ is the velocity spread of the pair plasma).
 The timescale for soliton collapse, and the duration of the
 nanoshot produced by each collapse, is therefore $\delta t \sim 1
 / \Lambda \nu_{\rm p}$. The spectrum of the nanoshot will be be relatively
 narrowband, $\delta \nu \sim 1 / \delta t \sim \Lambda \nu_{\rm p}$.
 Weatherall found that strong, localized solitons appear for
 $\Lambda \gtw 0.1$.

 Weatherall's calculation assumes the radiative loss time scale is no
 longer than dynamic timescale  for soliton
 collapse.  This holds for packet sizes smaller than $\sim c / \Lambda
 \nu_{\rm p}$, which is consistent with other estimates of soliton size
 (e.g., Weatherall \& Benford 1991).

Polarization of the nanoshot is another key test.  The collapsing
soliton generates electromagnetic waves with polarization set by
local plasma conditions.  Weatherall (1998) worked in the context of a
strongly magnetized plasma.  In this limit, emergent radiation around
the plasma frequency -- such as the nanoshots --  linearly
polarized along the magnetic field (c.f. the  O mode of Arons \& Barnard 1986).

Because strong Langmuir turbulence is an incoherent superposition of
localized solitons (e.g. Robinson 1997), the radiation
``shot'' produced by each soliton is independent of those from
its neighbors. This agrees nicely with the amplitude-modulated noise
model suggested by high-time-resolution observations of pulsar radio
emission (Rickett 1975, also Sec. \ref{MP_nanoshots}). We compare
 this model to the Crab  pulsar in  Sec. \ref{SPT}.

\subsection{Free-electron maser emission}
\label{App:FEM}

In this model, a relativistic beam encounters an electric field that
varies in space and time. Charges interacting with the field bunch
together in a resonant interaction, and emit intense, coherent bursts
of radiation.  This process has been extensively studied in the lab,
where particle beams pass through externally applied electromagnetic
fields (``wigglers'') and radiate in a narrow
cone beamed along the initial beam direction.

This interesting idea has not  been taken very far
in pulsar applications.  Fung \& Kuijpers (2004) proposed a model in which
a high-$\gamma$ beam propagates through a strong, transverse EM
wiggler in a nearly unmagnetized background plasma. 
 Their simulations found the expected
collective particle bunching, resulting in short-lived, narrow-band bursts
of radiation, on a timescale governed by the frequency of the imposed EM
wiggler. However, neither an externally applied transverse wiggler,
nor a nearly unmagnetized plasma, seem easy to maintain in a pulsar
magnetosphere.

A more likely possibility is that the particle beam generates its own
parallel wiggler.  Baker \etal (1988), also Weatherall
\& Benford (1991), suggested that when beam-driven Langmuir turbulence
 becomes strong, both it and the
driving beam will become spatially inhomogneous.  In that situation, 
 the collective motion of charge bunches interacting with electrostatic
fluctuations in the turbulence 
results in coherent, forward-beamed radiation (e.g. Kato \etal 1983,
Schopper \etal 2003).  For weak Langmuir turbulence, these
 models predict linearly polarized radiation
at $\sim \gamma^2 \nu_{\rm p}$, where $\nu_{\rm p}$ is the plasma frequency of the
background plasma. If turbulence
is strong enough to collapse to solitons with scale $D < c / \omp$, the
Compton-boosted beam radiation comes out at even higher frequencies
 $\sim \g_{\rm b}^2 c/D$ (Weatherall \& Benford 1991). In either case, the
radiation frequency is significantly boosted relative
to the plasma frequency of the background plasma.  We consider this for
the Crab pulsar in Sec. \ref{FEM}.

\subsection{Cycloton instability emission} 
\label{App:CIE}

In this model, a relativistic particle beam 
drives a cyclotron instability through the 
cyclotron resonance,  $\omega - k_{\parallel} v_{\parallel} - s \omega_B
/ \gamma = 0$.  Initially the $s = 1$ resonance was
invoked as a source of Alfv\'en waves that could scatter particles to higher
pitch angles as they passed into the wind, thus enabling  
synchrotron radiation (Machabeli \& Usov 1979) from the wind nebula.  

Later authors, including Kazbegi \etal (1991) and Lyutikov \etal (1999),
suggested the anomalous cyclotron 
instability ($s < 0 $) could be a direct radiation source,
generating transverse waves which  escape the plasma 
 without needing mode conversion. 
This interaction causes a beam particle to increase 
its perpendicular momentum, at the same time that it emits a photon; 
the energy of course comes from the parallel momentum. 
 If wave modes
in the background plasma are circularly polarized -- as is assumed in
their formulation -- the emitted radiation is also 
circularly polarized.  

In practice, this model is challenged in
application to real pulsars, because  of the 
weak magnetic fields and high plasma densities
 needed in order to put the emitted radiation in the radio
band. It is also challenged by the lack of a clear path to produce
short-lived nanoshots.  
 We consider this model for the Crab pulsar in  Sec. \ref{CIE}.

\subsection{Linear acceleration emission}
\label{App:LAE}

In this model, a beam of relativistic charges moving along a magnetic
field passes through and responds to an oscillating, parallel electric field.
The particles radiate due to the acceleration caused by $E_{\parallel}$. 
Melrose (1978) treated this process in the linear limit (assuming
$\gamma$ stays nearly constant), and showed that the
 radiation comes out around $\gamma^2 \omega_o$, 
where $\omega_o$ is the oscillation frequency of the E field. The 
radiation is linearly polarized parallel to the projected B
field. To reach high 
brightness temperatures while retaining the single-particle
linear analysis,  Melrose 
suggested negative absorption (maser amplification), but 
found the optical depth is  much less than unity for typical pulsar
conditions. 

The origin of the oscillating E field is unclear.  Melrose (1978)
suggested oscillation at $\sim \omp$, a case which is 
very similar to free-electron maser models when the wigglers are
self-generated Langmuir waves. Unsteady pair creation in an unshielded
gap has also been suggested as the  source of oscillating E fields 
(e.g. Levinson \etal 2005,  Timohkin \& Arons 2013).  These 
models are still being developed, and the time signature of the unshielded
E field  is not yet clear. In addition, recent
 work has explored electric fields strong enough to 
change the particle's Lorentz factor significantly in one oscillation. 
This limit seems more relevant to
high-enegy emission, and different treatments are still being explored 
in the literature (Melrose \etal  2009, Reville \& Kirk 2010).  We find this
mechanism interesting, but further 
work is needed before it can be usefully tested against pulsar
data.

\subsection{Maser amplification of radio beams}
\label{App:masers}

Lyubarskii \& Petrova (1996) considered stimulated Compton scattering of
pulsar radio emission
by the magnetospheric pair plasma in the presence of background radiation.
They were 
interested in scattering out of the radio beam, to explain
low-frequency spectral turnovers seen in many pulsars.  However in
later papers, Petrova (e.g., 2008, 2009) suggested that stimulated
scattering of radio emission out of the primary beam can be the origin
of secondary components in a pulsar's mean profile (e.g. the
interpulse found in a few radio pulsars, or the several components of
the Crab's mean profile).  Unfortunately, the scattering opacity is hard
to predict.   It depends not just on magnetospheric parameters,
but also varies with scattering regimes,  so that the nature
of the scattered radiation is extremely sensitive to assumed details of
the scattering.  While these models have some nice
physics,  the difficulty of quantifying the scattering opacity 
seems to preclude any robust test of them against the data.

Weatherall (2001) also considered stimulated Compton scattering of a
pulsar radio beam, but  by strong turbulence in the magnetospheric
plasma which the beam traverses.  He mostly studied scattering in a
nonmagnetized plasma, but did include a simple model restricting charge
motion to one dimension
 (appropriate for a high magnetic field).  He finds that interesting
amplification is possible if the plasma turbulence extremely strong
(e.g. turbulent energy density on the order of $10^{12}$ times the
plasma temperature).  However, he presents no discussion of how such strong
turbulence can be maintained, nor does he consider radiation emitted 
by the turbulence itself (as in Sec. \ref{App:SPT} above).  

In addition, both of these models assume the existence of an initial
radio beam that can be amplified by stimulated scattering. 
The origin of the  beam is not discussed.
In that sense, neither of these models addresses  the primary
radio emission mechanism. Given the many uncertainties,  neither model can
easily be compared to radio data. We do not pursue these any further.

\subsection{Coherent charge bunches} 
\label{App:ChargeBunches}

Curvature emission by coherent bunches was one of the first models
proposed for pulsar radio emission (e.g. Ruderman \& Sutherland
1975, Buschauer \& Benford 1976).  The general idea is that some
instability groups $N \gg 1$ charges into a coherently moving bunch
that acts as a single large charge, $Q = N e$; this bunch then
radiates curvature emission in the radio band. Although this model
initially seemed attractive, subsequent analysis (e.g., Melrose 1978,
1995, and references therein) identified several serious problems,
including but not limited to the lack of an effective mechanism for
forming the bunches, and the inability of the putative charge bunch to
hold together long enough to radiate.

This model has recently been revived by Melikidze \etal (2000; also Gil
\etal 2004).  They follow Karpman \etal (1975), who argued that Langmuir
solitons will be charge-separated if the two charge species have
different masses.  Melikidze \etal (2000) proposed that solitons in a
pair plasma will also be charge-separated if the electrons and
positrons have different relativistic streaming speeds.  These authors
then assumed that the solitons are sufficiently long-lived for the
separated ``charges'' within each soliton to produce curvature
emission.  They further assumed that the plasma and soliton parameters
are just what is needed to put the radiation in the radio band, and they
ignored any other radiation caused by plasma dynamics within the soliton.  
Given the large number of assumptions required by this model, and the
lack of any demonstration of the solitons' dynamic evolution,
we do not consider this model any further. 

\subsection{ Curvature maser emission}
\label{App:CurvDrift}

While simple curvature emission cannot support maser action (Blandford
1975), signal amplification is possible if 
particle curvature drift and/or field line distortion is taken into
account (e.g. Luo \& Melrose 1992, Luo \& Melrose 1994, Lyutikov \etal
1999). However, the model details are very sensitive to the local field line
curvature, which is  poorly known in the outer magnetosphere.  
The situation is also complicated by the fact that the energetic particle beam
necessary for either type of curvature maser also 
drives streaming instabilities that
can lead to relativistic plasma emission (see Melrose 1995). 
Furthermore, there is still disagreement in the literature about the
efficacy of this mechanism (e.g., Kaganovich \& Lyubarsky 2010).  For all
of these reasons, we do,  not pursue this mechanism here.

\section{Double plasma resonance model for pulsar emission}
\label{App:DPR}

The spectral emission bands in the High-Frequency Interpulse have not been
seen in any other pulsar.  A variety of new
models, specific to the emission bands, have been suggested since
their discovery, but none have successfully explained all the data.
 We review several such models in Hankins \etal (2016); here we only
discuss one, the Double Plasma Resonance 
which may explain the ``zebra'' emission bands
seen in Type IV solar flares (e.g., Kuijpers 1975, Zhelezniakov \& Zlotnik
1975, Winglee \& Dulk 1986).

This is another two-stage process.  A maser
instability generates upper hybrid waves at the cyclotron resonance, and those
waves  must be converted to transverse modes in order to escape the system.  
The instability growth is especially rapid at a set 
of discrete frequencies, 
\be
\nu_{\rm res} 
= \left( \nu_{\rm p}^2 + \nu_{\rm  B}^2 \right)^{1/2} = s \nu_{\rm B}
\label{DPR_full}
\ee
 where
$s$ is the harmonic number, $\nu_{\rm p}$ is the plasma frequency,
 $\nu_{\rm B}$ is the electron
cyclotron frequency, and the growth is fastest for waves propagating
{\it across} the local magnetic field.

 If the plasma is non-uniform, with the magnetic
field and plasma density varying on different scales, the resonance
for each harmonic $s$ will be satisfied at a different spatial location,
and multiple spectral bands will be seen.  This is a stringent constraint.  
In order to produce a particular value of the band spacing, $\Delta
\nu / \nu$, the local structure of the plasma density and magnetic
field must have just the right behavior.  From equation (\ref{DPR_full}),
in the limit $\nu_{\rm B} \ll \nu_{\rm p}$, the
spacing between adjacent bands is usually written
\be
{\Delta \nu \over \nu} \simeq { L_{\rm B} \over  s L_{\rm B} - (s+1) L_{\rm n}}
\label{DPR_grads}
\ee
where $L_{\rm B}$ and $L_{\rm n}$
 are the gradient scales of the magnetic field and
density:  $L_{\rm B} = B / |\nabla B|$ and  $L_{\rm n} = 2 n/| \nabla n| $. 
 For general
field and density distributions, $L_{\rm B}$ and $L_{\rm n}$ are 
functions of position,
and equation (\ref{DPR_grads}) becomes an implicit condition for the 
location
at which equation (\ref{DPR_full}) is satisfied for a given harmonic $s$. 

In addition to the geometrical restriction, this theory has other
important requirements on the plasma.  The magnetic field must be
weak, $\nu_{\rm B} \ll \nu_{\rm p}$, and much lower than is predicted anywhere in
the standard magnetosphere if $\nu_{\rm res}$ is to lie in the radio band.
  The theory has only been developed for
non-relativistic, electron-ion plasmas. Zheleznyakov \etal (2012)
suggested the instability may also develop in pair plasmas, but we have not
found that work in the literature.  There must be an excess of
energetic particles moving across the magnetic field -- a
situation which is common in solar coronal loops, but not expected in
the standard magnetosphere.

Despite the challenges, we 
find this model attractive, because it offers a clear physical cause of
discrete spectral bands such as those we see in the High-Frequency Interpulse
of the Crab pulsar. We discuss its relevance for the Crab pulsar in Sec.
\ref{EmissBandModels}.

\bibliographystyle{jpp}
%\bibliography{jpp-instructions}

\begin{thebibliography}{85}  %Note need explicit count of number of ref's
\expandafter\ifx\csname natexlab\endcsname\relax\def\natexlab#1{#1}\fi

\bibitem[Abdo \etal (2010)]{Abdo2010}
{  Abdo, A.~A., Ackerman, M., Ajello, M. \etal} 2010 Fermi Large Area 
Telescope observations of the Crab pulsar and Nebula. \ApJ {\bf 708},
 1254--1267.

\bibitem[Abdo \etal (2013)]{Abdo2013}
{  Abdo, A.~A., Ajello, M., Allafort \etal} 2013 The second Fermi
Large Area Telescope catalog of gamma-ray pulsars. \ApJSupp {\bf 208},
 17.

\bibitem[Allen \& Melrose (1982)]{AM82}
{  Allen, M. C. \& Melrose, D. B.} 1982 Elliptically polarized natural modes
in pulsar magnetospheres. {\it Proc. Ast. Soc. of Australia}, {\bf 4}, 365--370.

\bibitem[Arendt \& Eilek (2002)]{AE02}
{  Arendt, P. N. Jr. \& Eilek, J. A.} 2002 Pair creation in the pulsar
magnetosphere. \ApJ {\bf 581}, 451--469.

\bibitem[Arons \& Barnard (1986)]{AB86}
{  Arons, J. \& Barnard, J. J.} 1986 Wave propagation in pulsar 
magnetospheres -- dispersion relations and normal modes of plasmas in 
superstrong magnetic fields. \ApJ {\bf 302} 120--137.

\bibitem[Arons \& Scharlemann (1979)]{AS79}
{  Arons, J. \& Scharlemann, E. T.} 1979 Pair formation above pulsar 
polar caps -- structure of the low altitude acceleration zone. \ApJ {\bf 231},
854--879.

\bibitem[Bai \& Spitkovsky (2010)]{BaiS2010}
{  Bai, X.-N. \& Spitkovsky, A.} 2010 Modeling of gamma-ray pulsar light
curves using the force-free magnetic field. \ApJ {\bf 715}, 1282--1301.

\bibitem[Baker \etal (1988)]{Bak88}
{  Baker, D. N., Borovsky, J. E., Benford, G. \& Eilek, J. A.} 1988 The
collective emission of electromagnetic waves from astrophysical jets:
luminosity gaps, BL Lacertae objects, and efficient energy transport. \ApJ 
{\bf 326}, 110--124.

\bibitem[Benford (1992)]{Ben92}
{  Benford, G.} 1992 Broadband microwave generation by beam-plasma
turbulence. {\it IEEE Trans. Plasma Sci.} {\bf 20}, 370--372.


\bibitem[Benford \& Tzach (2000)]{BT00}
{  Benford, G. \& Tzach, D.} 2000 Coherent synchroton emission observed:
implications for radio astronomy. \MNRAS {\bf 317}, 497--500.

\bibitem[Blandford (1975)]{Bland75}
{  Blandford, R. D.} 1975 Amplification of radiation by relativistic particles
in a strong magnetic field. \ApJ {\bf 170}, 551--557.

\bibitem[Bogovalov(1999)]{BO99}
{  Bogovalov, S. V.} 1999 On the physics of cold MHD winds from oblique 
rotators. \AandA {\bf 349}, 1017--1026.

\bibitem[Bucciantini \etal (2010)]{BAA10}
{  Bucciantini, N., Arons, J. \& Amato, E.} 2010 Modelling spectral evolution
of pulsar wind nebulae inside supernova remnants. \MNRAS {\bf 410} 381--389.

\bibitem[Buschauer \& Benford (1976)]{BB76}
{  Buschauer, R. \& Benford, G.} 1976 General theory of coherent curvature
radiation. \MNRAS {\bf 177}, 109--136.

\bibitem[Carlqvist (1982)]{Car82}
{  Carlqvist, P.} 1982 On the physics of relativistic double layers. {\it
Astrophys. \& Space Sci.} {\bf 87}, 21--39.

\bibitem[Cerutti \etal (2015)]{Cer15}
{  Cerutti, B., Philippov, A., Parfrey, K. \& Spitkovsky, A.} 2015 Particle
acceleration in axisymmetric pulsar current sheets. \MNRAS {\bf 448}, 606--619.

\bibitem[Chen \etal (2011)]{Chen11}
{  Chen, B., Bastian. T. S., Gary, D. E. \& Jing, J.} 2011 Spatially and
spectrally resolved observations of a zebra pattern in a solar decimetric radio
burst. \ApJ {\bf 736}, 64. 

\bibitem[Cheng \etal (1986)]{CHR86}
{  Cheng, K. S., Ho, D. \& Ruderman, M.} 1986 Energetic radiation from 
rapidly spinning pulsars.  I -- outer magnetosphere gaps. \ApJ {\bf 300},
500--521.

\bibitem[Cheng \etal (2000)]{CHZ00}
{  Cheng, K. S.,  Ruderman, M. \& Zhang, L.} 2000 A three-dimensional
outer magnetospheric gap model for gamma-ray pulsars:  geometry, pair
production, emission morphologies, and phase-resolved spectra.  \ApJ {\bf 537},
964--976.

\bibitem[Chernov \etal (2005)]{Cher05}
{  Chernov, G. P., Yan, Y. H., Fu, Q. J. \& Tan, Ch. M.} 2005 Recent data on
zebra patterns. \AandA {\bf 437}, 1047--1054.

\bibitem[Contopoulos \& Kalapotharakos (2010)]{CK10}
{  Contopoulos, I. \& Kalapotharakos, C.} 2010 The pulsar synchrotron in
3D:  curvature radiation. \MNRAS {\bf 404}, 767--778.

\bibitem[Crossley et al (2010)]{Crossley10}
{  Crossley, J.~H., Eilek, J.~A., Hankins, T.~H. \& Kern, J.~S.} 
2010 Short-lived
radio bursts from the Crab pulsar. \ApJ {\bf 722}, 1908--1920.

\bibitem[Deutsch (1955)]{Deutsch}
{  Deutsch, A.~J.} 1955 The electromagnetic field of an idealized star in 
rigid rotation in vacuo. {\em  Annales d'Astrophysique}, {\bf 18}, 1--10.

\bibitem[Dyks \etal (2004)]{Dyks04}
{  Dyks, J., Harding, A.~K. \& Rudak, B.} 2004 Relativistic effects and
polarization in three high-energy pulsar models. \ApJ {\bf 606}, 1125--1142.

\bibitem[Fung \& Kuijpers (2004)]{FK04}
{  Fung, P. K. \& Kuijpers, J.} 2002 A free-electron laser in the pulsar
magnetosphere. \AandA {\bf 422}, 817--830.

\bibitem[Gil \etal (2004)]{Gil04}
{  Gil, J., Lyubarsky, Y. \& Melikidze, G. I.} 2004 Curvature radiation in
pulsar magnetospheric plasma. \ApJ {\bf 600}, 872--882.

\bibitem[Goldreich \& Julian (1969)]{GJ69}
{  Goldreich, P. \& Julian, W. H.} 1969 Pulsar electrodynamics.
\ApJ {\bf 157}, 869--880.

\bibitem[Hankins et al (2003)]{HKWE03}
{  Hankins, T.~H., Kern, J.~S., Weatherall, J.~C. \& Eilek, J.~A.} 2003
Microsecond radio bursts from strong plasma turbulence in the Crab pulsar.
{\em Nature} {\bf 422}, 141--143.

\bibitem[Hankins \& Eilek  (2007)]{HE07}
{  Hankins, T.~H.  \& Eilek, J.~A.} 2007 Radio emission signatures in 
the Crab pulsar. \ApJ {\bf 670}, 693--701.

\bibitem[Hankins et al  (2015)]{HJE2015}
{  Hankins, T.~H., Jones, G.  \& Eilek, J.~A.} 2015 The Crab pulsar at 
centimeter wavelengths. 1.  Ensemble Characteristics. \ApJ {\bf 802}, 130.

\bibitem[Hankins et al  (2016)]{HJE2016}
{  Hankins, T.~H., Eilek, J.~A. \& Jones, G.} 2016 The Crab pulsar at 
centimeter wavelengths. 2.  Single pulse characteristics.  Submitted to \ApJ

\bibitem[Hibschman  \& Arons (2001)]{HA01}
{  Hibschman, J.~A. \& Arons, J.} 2001 Pair production multiplicities in
rotation-powered pulsars. \ApJ {\bf 560}, 871--884.

\bibitem[Ho (1989)]{H89}
{  Ho, C.} 1989 Spectra of Crab-like pulsars. \ApJ {\bf 342}, 396--405.

\bibitem[Kaganovich \& Lyubarsky (2010)]{KL10}
{  Kaganovich, A. \& Lyubarsky, Y.} 2010 Curvature-drift instability fails
to generate pulsar radio emission. \ApJ {\bf 721}, 1164--1173.

\bibitem[Kalapotharakos \etal (2012a)]{Kal12a}
{  Kalapotharakos, C., Contopoulos, I. \& Kazanas, D.}, 2012a The extended
pulsar magnetosphere. \MNRAS, {\bf 420}, 2793--2798.

\bibitem[Kalapotharakos \etal (2012b)]{Kal12b}
{  Kalapotharakos, Kazanas, D., Harding, A. \&  Contopoulos, I.} 2012b 
Toward a realistic pulsar magnetosphere. \ApJ {\bf 749}, 2.

\bibitem[Karpman \etal (1975)]{Ketal75}
{  Karpman, V. I., Norman, C. A., Ter Harr, D. \& Tsytovich, V. N.} 1975
Relativistic solitons and pulsars. {\it Phys. Scripta} {\bf 11}, 271--274.

\bibitem[Kato \etal (1983)]{Kato83}
{  Kato, K. G., Benford, G. \& Tzach, D.} 1983 Detailed spectra of high-power
broadband microwave radiation from interactions of relativistic electron
beams with weakly magnetized plasmas. {\em Physics of Fluids}
{\bf 26}, 3636--3649.

\bibitem[Kazbegi \etal (1991)]{Kaz91}
{  Kazbegi, A. K., Machabeli, G. Z. \& Melikidze, G. I.} 1991 On the
circular polarization in pulsar emission. \MNRAS {\bf 253}, 377--387.

\bibitem[Kuijpers (1975)]{Kuip75}
{  Kuijpers J.} 1975 A unified explanation of solar type IV DM continua and
ZEBRA patterns. \AandA {\bf 40}, 405--410.

\bibitem[Kuijpers (1990)]{Kuip90}
{  Kuijpers, J.} 1990 Coherent radiation from electrostatic double layers,
in 
{\it Plasma Phenomena in the Solar Atmosphere}, ed. M. Dubois, F. Bely-Dubau
\& D. Gr\`esellon (Les Editions de Physique, France) p. 17.

\bibitem[Kunzl \etal (1998)]{K98}
{  Kunzl, T., Lesch, H., Jessner, A. \& von Hoensbroech, A.} 1998 On pair 
production in the Crab pulsar.  \ApJ {\bf 505}, L139--L141.

\bibitem[Labelle \etal (2003)]{Lab03}
{  LaBelle, J., Truemann, R. A., Yoon, P. H. \& Karlick\'y, M.} 2003 A model
of zebra emission in solar Type IV radio bursts. \ApJ {\bf 593}, 1195--1207.

\bibitem[Ledenev \etal (2001)]{Led01}
{  Ledendev, V. G., Karlick\' y, M., Yan, Y. \& Fu, Q.} An estimation of
the coronal magnetic field strength from spectrographic observations in the
microwave range. {\it Solar Phys.} {\bf 202}, 71--79.

\bibitem[Levinson \etal (2005)]{Lev05}
{  Levinson, A., Melrose, D., Judge, A. \& Luo, Q.} 2005 Large-amplitude,
pair-creating oscillations in pulsar and black hole magnetospheres. \ApJ 
{\bf 631}, 456--465.

\bibitem[Li \etal (2012)]{Li12}
{  Li, J., Spitkovsky, A. \& Tchekhovskoy, A.} 2012 Resistive solutions for
pulsar magnetospheres. \ApJ {\bf 746}, 60.

\bibitem[Luo \& Melrose (1992)]{LM92}
{  Luo, Q. \& Melrose, D. B.} 1992 Coherent curvature emission and radio
pulsars. \MNRAS {\bf 258}, 616--620.

\bibitem[Luo \& Melrose (1995)]{LM95}
{  Luo, Q. \& Melrose, D. B.} 1995 Curvature maser emission due to field line
torsion in pulsar magnetospheres. \MNRAS {\bf 276}, 372--382.

\bibitem[Lyubarskii \& Petrova (1996)]{LB96}
{  Lyubarskii, Yu. E. \& Petrova, S. A.} 1996 Stimulated scattering of
radio emission in pulsar magnetospheres. {\em Astr. Lett.\/}, {\bf 22}, 
399--408.

\bibitem[Lyutikov \etal (1999)]{Lyu99}
{  Lyutikov, M., Machabeli, G. \& Blandford, R.} 1999 Cherenkov-curvature 
radiation and pulsar radio emission generation. \ApJ {\bf 512}, 804--826.

\bibitem[Machabeli \& Usov(1979)]{MachUsov79}
{  Machabeli, G. Z. \& Usov, V. V.} 1979 Cyclotron instability in the
magnetosphere of the Crab Nebula pulsar, and the origin of its radiation.
{\it Soviet Ast. Lett.} {\bf 5}, 238--241.

\bibitem[Melikidze \etal (2000)]{Metal00}
{  Melikidze, G. I., Gil, J. A. \& Pataraya, A. D.} 2000 The spark-associated
soliton model for pulsar radio emission. \ApJ {\bf 544}, 1081--1096.

\bibitem[Melrose (1978)]{Mel78}
{  Melrose, D. B.} 1978 Amplified linear acceleration emission applied to
pulsars. \ApJ {\bf 225}, 557--573.

\bibitem[Melrose (1995)]{Mel95}
{  Melrose, D. B.} 1995 The models for radio emission from pulsars -- the 
outstanding issues. {\em J. Astrophys. Astr.} {\bf 16}, 137--164.

\bibitem[Melrose \etal (2009)]{Mel09}
{  Melrose, D. B., Rafat, M. Z. \& Luo, Q.} 2009 Linear acceleration 
emission I:  motion in a large-amplitude electrostatic wave. \ApJ 
{\bf 698}, 115--123.

\bibitem[Melrose \& Gedalin (1999)]{MG99}
{  Melrose, D. B. \& Gedalin, M. E.} 1999 Relativistic plasma emission
and pulsar radio emission:  a critique. \ApJ {\bf 521}, 351--361.

\bibitem[Moffett \& Hankins (1996)]{MH96}
{  Moffett, D. A. \& Hankins, T.~H.} 1996 Multifrequency radio observations
of the Crab pulsar.  \ApJ {\bf 468}, 779--783.

\bibitem[Moffett \& Hankins (1999)]{MH99}
{  Moffett, D. A. \& Hankins, T.~H.} 1999 Polarimetric properties of
the Crab pulsar between 1.4 and 8.4 GHz.  \ApJ {\bf 522}, 1046--1052.

\bibitem[Muslimov \& Harding (2004)]{MH04}
{  Muslimov, A. G. \& Harding, A. K.} High-altitude particle acceleration 
and radiation in slot gaps. \ApJ {\bf 606}, 1143--1153.

\bibitem[Ng \& Romani (2004)]{NgRom2004}
{  Ng, C.-Y. \& Romani, R.~W.} 2004 Fitting pulsar wind tori.
 \ApJ {\bf 601}, 479--484.

\bibitem[Qiao \etal (2004)]{Qiao04}
{  Qiao, G. J., Lee, K. J., Wang, H. G. \etal} 2004 The inner annular gap for
pulsar radiation:  gamma-ray and radio emission. \ApJ {\bf 606} L49--L52.

\bibitem[Petri \& Kirk (2005)]{PK05}
{  P\'etri, J. \& Kirk, J. G.}, 2005 The polarization of high-energy pulsar 
radiation in the striped wind model. \ApJ {\bf 627}, L37--L40.

\bibitem[Petrova (2008)]{Pet08}
{  Petrova, S. A.} 2008 Interpretation of the low-frequency peculiarities in
the radio profile structure of the Crab pulsar. \MNRAS {\bf 385}, 2143--2150.

\bibitem[Petrova (2009)]{Pet09}
{  Petrova, S. A.} 2009 Formation of the radio profile components of the
Crab pulsar. \MNRAS {\bf 395}, 1723--1732.

\bibitem[Radhakrishnan  \& Cooke (1969)]{RadC69}
{  Radhakrishnan, V. \& Cooke, D. J.} 1969 Magnetic poles and the
polarization structure of pulsar radiation. {\em Astrophys Lett.}, {\bf 3},
225--229.

\bibitem[Rankin \etal (1970)]{Rankin1970}
{  Rankin, J.~M., Comella, J.~M., Craft, H.~D.~Jr. \etal}
1970 Radio pulse shapes, flux  densities,
and dispersion of pulsar NP 0532.  \ApJ {\bf 162}, 707--725.

\bibitem[Rickett(1975)]{Rickett75}
{  Rickett, B.~J.} 1975 Amplitude-modulated noise:  an empirical model 
for the radio radiation received from pulsars. \ApJ {\bf 197}, 185--191.

\bibitem[Reville \& Kirk (2010)]{RK10}
{  Reville, B. \& Kirk, J. G.} 2010 Linear acceleration emission in
pulsar magnetospheres. \ApJ {\bf 715}, 186--193.

\bibitem[Robinson(1997)]{Rob77}
{  Robinson, P. A.} 1977 Nonlinear wave collapse and strong turbulence.
{\em Rev. Mod. Phys.\/} {\bf 69}, 507--572.

\bibitem[Romani \& Yadigaroglu (1995)]{RomY95}
{  Romani, R. W. \& Yadigaroglu, I.-A.} 1995 Gamma-ray pulsars: 
 emission zones and viewing geometries. \ApJ {\bf 438} 314--321.

\bibitem[Ruderman \& Sutherland (1975)]{RS75}
{  Ruderman, M. A. \& Sutherland, P. G.} 1975 Theory of Pulsars -- Polar caps,
sparks, and coherent microwave radiation.  \ApJ {\bf 196}, 51--72.

\bibitem[Schopper \etal (2003)]{Sch03}
{  Schopper, R., Ruhl, H., Kunzl, T. A. \& Lesch, H.} 2003 Kinetic simulation
of the coherent radio emission from pulsars. {\em Laser and Particle Beams} 
{\bf 21}, 109--113.

\bibitem[Shearer \etal (2012)]{Sh12}
{  Shearer, A., Collins, S., Naletto, G. \etal } 2012 High-time-resolution
optical observations of the Crab pulsar; in 
 {\em Electromagnetic radiation from  pulsars and magnetars} ed. W.
Lewandoski, O. Maron, J. Kijak \& A. S\l owikowska
 (San Francisco: Astronomical Society of the Pacific), 11--14.

\bibitem[Shcherbakov (2008)]{Shch08}
{  Shcherbakov, R. V.} 2008 Propagation effects in magnetized transrelativistic
plasmas. \ApJ {\bf 688}, 695--700.


\bibitem[S\l owikowska \etal (2009)]{Slow09}
{  S\l owikowska, A., Kanbach, G., Kramer, M. \& Stefanescu, A.} 2009 Optical
polarization of the Crab pulsar:  precision measurements and comparison to the
radio emission. \MNRAS {\bf 397}, 103--123.

\bibitem[Timokhin \& Arons (2013)]{TA13}
{  Timokhin, A. N. \& Arons, J.} 2013 Current flow and pair creation at low
altitude in rotation-powered pulsars' force-free magnetospheres:  space
charge limited flow. \MNRAS {\bf 429}, 20--54.

\bibitem[Weatherall(1997)]{JCW97}
{  Weatherall, J. C.} 1997 Modulational instability, mode conversion and radio
emission in the magnetized pair plasma of pulsars. \ApJ {\bf 483}, 402--413.

\bibitem[Weatherall(1998)]{JCW98}
{  Weatherall, J.~C.} 1998 Pulsar radio emission by conversion of plasma
wave turbulence:  nanosecond time structure. \ApJ {\bf 506}, 341--346.

\bibitem[Weatherall(2001)]{JCW01}
{  Weatherall, J.~C.} 2001 A relativistic-plasma Compton maser. \ApJ
{\bf 559}, 196--200.

\bibitem[Weatherall \& Benford(1991)]{WB91}
{  Weatherall, J. C. \& Benford, G.} 1991 Coherent radiation from energetic
electron streams via collisionless bremsstrahlung in strong plasma turbulence.
\ApJ {\bf 378}, 543--549.

\bibitem[Windsor \& Kellog (1974)]{WK74}
{  Windsor, R. A. \& Kellog, P. J.} 1974 Polarization of inverse plasmon
scattering. \ApJ {\bf 190}, 167--173.

\bibitem[Winglee \& Dulk (1986)]{WD86}
{  Winglee, R. M. \& Dulk, G. A.} 1986 The electron-cyclotron maser instability
as a source of plasma radiation. \ApJ {\bf 307}, 808--819.

\bibitem[Zampieri \etal (2014)]{Zampieri2014}
{  Zampieri, L., Cadez, A., Barbiere, C. et al.} 2014 Optical phase coherent
timing of the Crab nebula pulsar with Iqueye at the ESO New Technology
Telescope. \MNRAS {\bf 439}, 2813--2821.

\bibitem[Zheleznakov \& Zlotnik (1975)]{ZZ75}
{  Zhelezniakov, V. V. \& Zlotnik, E. Ia.} 1975 Cyclotron wave instability
in the corona and origin of solar radio emission with fine structure.  
I -- Bernstein modes and plasma waves in a hybrid band. {\it Solar Phys.} 
{\bf 43}, 431--451.

\bibitem[Zheleznakov \etal (2012)]{Zetal12}
{  Zheleznyakov, V. V., Zaitsev, V. V. \& Zlotnik, E. Ya.} 2012 On the
analogy between the zebra patterns in radio emission from the sun and
the Crab pulsar. {\it Ast. Lett.} {\bf 9} 589--604.


\end{thebibliography}

\end{document}